\magnification=\magstep1
\def\firstpage{1}
\pageno=\firstpage
\font\fiverm=cmr5
\font\sevenrm=cmr7
\font\sevenbf=cmbx7
\font\eightrm=cmr8
\font\eightbf=cmbx8
\font\ninerm=cmr9
\font\ninebf=cmbx9
\font\tenbf=cmbx10
\font\twelvebf=cmbx12
%

%

%

%
%
\newskip\ttglue
\font\fiverm=cmr5
\font\fivei=cmmi5
\font\fivesy=cmsy5
\font\fivebf=cmbx5
\font\sixrm=cmr6
\font\sixi=cmmi6
\font\sixsy=cmsy6
\font\sixbf=cmbx6
\font\sevenrm=cmr7
\font\eightrm=cmr8
\font\eighti=cmmi8
\font\eightsy=cmsy8
\font\eightit=cmti8
\font\eightsl=cmsl8
\font\eighttt=cmtt8
\font\eightbf=cmbx8
\font\ninerm=cmr9
\font\ninei=cmmi9
\font\ninesy=cmsy9
\font\nineit=cmti9
\font\ninesl=cmsl9
\font\ninett=cmtt9
\font\ninebf=cmbx9
\font\twelverm=cmr12
\font\twelvei=cmmi12
\font\twelvesy=cmsy12
\font\twelveit=cmti12
\font\twelvesl=cmsl12
\font\twelvett=cmtt12
\font\twelvebf=cmbx12


\def\eightpoint{\def\rm{\fam0\eightrm}
  \textfont0=\eightrm \scriptfont0=\sixrm \scriptscriptfont0=\fiverm
  \textfont1=\eighti  \scriptfont1=\sixi  \scriptscriptfont1=\fivei
  \textfont2=\eightsy  \scriptfont2=\sixsy  \scriptscriptfont2=\fivesy
  \textfont3=\tenex  \scriptfont3=\tenex  \scriptscriptfont3=\tenex
  \textfont\itfam=\eightit  \def\it{\fam\itfam\eightit}
  \textfont\slfam=\eightsl  \def\sl{\fam\slfam\eightsl}
  \textfont\ttfam=\eighttt  \def\tt{\fam\ttfam\eighttt}
  \textfont\bffam=\eightbf  \scriptfont\bffam=\sixbf
    \scriptscriptfont\bffam=\fivebf  \def\bf{\fam\bffam\eightbf}
  \tt  \ttglue=.5em plus.25em minus.15em
  \normalbaselineskip=9pt
  \setbox\strutbox=\hbox{\vrule height7pt depth2pt width0pt}
  \let\sc=\sixrm  \let\big=\eightbig \normalbaselines\rm}

\def\eightbig#1{{\hbox{$\textfont0=\ninerm\textfont2=\ninesy
        \left#1\vbox to6.5pt{}\right.$}}}


\def\ninepoint{\def\rm{\fam0\ninerm}
  \textfont0=\ninerm \scriptfont0=\sixrm \scriptscriptfont0=\fiverm
  \textfont1=\ninei  \scriptfont1=\sixi  \scriptscriptfont1=\fivei
  \textfont2=\ninesy  \scriptfont2=\sixsy  \scriptscriptfont2=\fivesy
  \textfont3=\tenex  \scriptfont3=\tenex  \scriptscriptfont3=\tenex
  \textfont\itfam=\nineit  \def\it{\fam\itfam\nineit}
  \textfont\slfam=\ninesl  \def\sl{\fam\slfam\ninesl}
  \textfont\ttfam=\ninett  \def\tt{\fam\ttfam\ninett}
  \textfont\bffam=\ninebf  \scriptfont\bffam=\sixbf
    \scriptscriptfont\bffam=\fivebf  \def\bf{\fam\bffam\ninebf}
  \tt  \ttglue=.5em plus.25em minus.15em
  \normalbaselineskip=11pt
  \setbox\strutbox=\hbox{\vrule height8pt depth3pt width0pt}
  \let\sc=\sevenrm  \let\big=\ninebig \normalbaselines\rm}

\def\ninebig#1{{\hbox{$\textfont0=\tenrm\textfont2=\tensy
        \left#1\vbox to7.25pt{}\right.$}}}


\def\twelvepoint{\def\rm{\fam0\twelverm}
  \textfont0=\twelverm \scriptfont0=\eightrm \scriptscriptfont0=\sixrm
  \textfont1=\twelvei  \scriptfont1=\eighti  \scriptscriptfont1=\sixi
  \textfont2=\twelvesy  \scriptfont2=\eightsy  \scriptscriptfont2=\sixsy
  \textfont3=\tenex  \scriptfont3=\tenex  \scriptscriptfont3=\tenex
  \textfont\itfam=\twelveit  \def\it{\fam\itfam\twelveit}
  \textfont\slfam=\twelvesl  \def\sl{\fam\slfam\twelvesl}
  \textfont\ttfam=\twelvett  \def\tt{\fam\ttfam\twelvett}
  \textfont\bffam=\twelvebf  \scriptfont\bffam=\eightbf
    \scriptscriptfont\bffam=\sixbf  \def\bf{\fam\bffam\twelvebf}
  \tt  \ttglue=.5em plus.25em minus.15em
  \normalbaselineskip=11pt
  \setbox\strutbox=\hbox{\vrule height8pt depth3pt width0pt}
  \let\sc=\sevenrm  \let\big=\twelvebig \normalbaselines\rm}

\def\twelvebig#1{{\hbox{$\textfont0=\tenrm\textfont2=\tensy
        \left#1\vbox to7.25pt{}\right.$}}}
\catcode`\@=11
%

\def\undefine#1{\let#1\undefined}
\def\newsymbol#1#2#3#4#5{\let\next@\relax
 \ifnum#2=\@ne\let\next@\msafam@\else
 \ifnum#2=\tw@\let\next@\msbfam@\fi\fi
 \mathchardef#1="#3\next@#4#5}
\def\mathhexbox@#1#2#3{\relax
 \ifmmode\mathpalette{}{\m@th\mathchar"#1#2#3}%
 \else\leavevmode\hbox{$\m@th\mathchar"#1#2#3$}\fi}
\def\hexnumber@#1{\ifcase#1 0\or 1\or 2\or 3\or 4\or 5\or 6\or 7\or 8\or
 9\or A\or B\or C\or D\or E\or F\fi}

\font\tenmsa=msam10
\font\sevenmsa=msam7
\font\fivemsa=msam5
\newfam\msafam
\textfont\msafam=\tenmsa
\scriptfont\msafam=\sevenmsa
\scriptscriptfont\msafam=\fivemsa
\edef\msafam@{\hexnumber@\msafam}
\mathchardef\dabar@"0\msafam@39
\def\dashrightarrow{\mathrel{\dabar@\dabar@\mathchar"0\msafam@4B}}
\def\dashleftarrow{\mathrel{\mathchar"0\msafam@4C\dabar@\dabar@}}

\def\ulcorner{\delimiter"4\msafam@70\msafam@70 }
\def\urcorner{\delimiter"5\msafam@71\msafam@71 }
\def\llcorner{\delimiter"4\msafam@78\msafam@78 }
\def\lrcorner{\delimiter"5\msafam@79\msafam@79 }
\def\yen{{\mathhexbox@\msafam@55}}
\def\checkmark{{\mathhexbox@\msafam@58}}
\def\circledR{{\mathhexbox@\msafam@72}}
\def\maltese{{\mathhexbox@\msafam@7A}}

\font\tenmsb=msbm10
\font\sevenmsb=msbm7
\font\fivemsb=msbm5
\newfam\msbfam
\textfont\msbfam=\tenmsb
\scriptfont\msbfam=\sevenmsb
\scriptscriptfont\msbfam=\fivemsb
\edef\msbfam@{\hexnumber@\msbfam}
\def\Bbb#1{{\fam\msbfam\relax#1}}
\def\widehat#1{\setbox\z@\hbox{$\m@th#1$}%
 \ifdim\wd\z@>\tw@ em\mathaccent"0\msbfam@5B{#1}%
 \else\mathaccent"0362{#1}\fi}
\def\widetilde#1{\setbox\z@\hbox{$\m@th#1$}%
 \ifdim\wd\z@>\tw@ em\mathaccent"0\msbfam@5D{#1}%
 \else\mathaccent"0365{#1}\fi}
\font\teneufm=eufm10
\font\seveneufm=eufm7
\font\fiveeufm=eufm5
\newfam\eufmfam
\textfont\eufmfam=\teneufm
\scriptfont\eufmfam=\seveneufm
\scriptscriptfont\eufmfam=\fiveeufm

\catcode`\@=11
\newsymbol\boxdot 1200
\newsymbol\boxplus 1201
\newsymbol\boxtimes 1202
\newsymbol\square 1003
\newsymbol\blacksquare 1004
\newsymbol\centerdot 1205
\newsymbol\lozenge 1006
\newsymbol\blacklozenge 1007
\newsymbol\circlearrowright 1308
\newsymbol\circlearrowleft 1309
\undefine\rightleftharpoons
\newsymbol\rightleftharpoons 130A
\newsymbol\leftrightharpoons 130B
\newsymbol\boxminus 120C
\newsymbol\Vdash 130D
\newsymbol\Vvdash 130E
\newsymbol\vDash 130F
\newsymbol\twoheadrightarrow 1310
\newsymbol\twoheadleftarrow 1311
\newsymbol\leftleftarrows 1312
\newsymbol\rightrightarrows 1313
\newsymbol\upuparrows 1314
\newsymbol\downdownarrows 1315
\newsymbol\upharpoonright 1316
 
\newsymbol\downharpoonright 1317
\newsymbol\upharpoonleft 1318
\newsymbol\downharpoonleft 1319
\newsymbol\rightarrowtail 131A
\newsymbol\leftarrowtail 131B
\newsymbol\leftrightarrows 131C
\newsymbol\rightleftarrows 131D
\newsymbol\Lsh 131E
\newsymbol\Rsh 131F
\newsymbol\rightsquigarrow 1320
\newsymbol\leftrightsquigarrow 1321
\newsymbol\looparrowleft 1322
\newsymbol\looparrowright 1323
\newsymbol\circeq 1324
\newsymbol\succsim 1325
\newsymbol\gtrsim 1326
\newsymbol\gtrapprox 1327
\newsymbol\multimap 1328
\newsymbol\therefore 1329
\newsymbol\because 132A
\newsymbol\doteqdot 132B
 
\newsymbol\triangleq 132C
\newsymbol\precsim 132D
\newsymbol\lesssim 132E
\newsymbol\lessapprox 132F
\newsymbol\eqslantless 1330
\newsymbol\eqslantgtr 1331
\newsymbol\curlyeqprec 1332
\newsymbol\curlyeqsucc 1333
\newsymbol\preccurlyeq 1334
\newsymbol\leqq 1335
\newsymbol\leqslant 1336
\newsymbol\lessgtr 1337
\newsymbol\backprime 1038
\newsymbol\risingdotseq 133A
\newsymbol\fallingdotseq 133B
\newsymbol\succcurlyeq 133C
\newsymbol\geqq 133D
\newsymbol\geqslant 133E
\newsymbol\gtrless 133F
\newsymbol\sqsubset 1340
\newsymbol\sqsupset 1341
\newsymbol\vartriangleright 1342
\newsymbol\vartriangleleft 1343
\newsymbol\trianglerighteq 1344
\newsymbol\trianglelefteq 1345
\newsymbol\bigstar 1046
\newsymbol\between 1347
\newsymbol\blacktriangledown 1048
\newsymbol\blacktriangleright 1349
\newsymbol\blacktriangleleft 134A
\newsymbol\vartriangle 134D
\newsymbol\blacktriangle 104E
\newsymbol\triangledown 104F
\newsymbol\eqcirc 1350
\newsymbol\lesseqgtr 1351
\newsymbol\gtreqless 1352
\newsymbol\lesseqqgtr 1353
\newsymbol\gtreqqless 1354
\newsymbol\Rrightarrow 1356
\newsymbol\Lleftarrow 1357
\newsymbol\veebar 1259
\newsymbol\barwedge 125A
\newsymbol\doublebarwedge 125B
\undefine\angle
\newsymbol\angle 105C
\newsymbol\measuredangle 105D
\newsymbol\sphericalangle 105E
\newsymbol\varpropto 135F
\newsymbol\smallsmile 1360
\newsymbol\smallfrown 1361
\newsymbol\Subset 1362
\newsymbol\Supset 1363
\newsymbol\Cup 1264
 
\newsymbol\Cap 1265
 
\newsymbol\curlywedge 1266
\newsymbol\curlyvee 1267
\newsymbol\leftthreetimes 1268
\newsymbol\rightthreetimes 1269
\newsymbol\subseteqq 136A
\newsymbol\supseteqq 136B
\newsymbol\bumpeq 136C
\newsymbol\Bumpeq 136D
\newsymbol\lll 136E
 
\newsymbol\ggg 136F
 
\newsymbol\circledS 1073
\newsymbol\pitchfork 1374
\newsymbol\dotplus 1275
\newsymbol\backsim 1376
\newsymbol\backsimeq 1377
\newsymbol\complement 107B
\newsymbol\intercal 127C
\newsymbol\circledcirc 127D
\newsymbol\circledast 127E
\newsymbol\circleddash 127F
\newsymbol\lvertneqq 2300
\newsymbol\gvertneqq 2301
\newsymbol\nleq 2302
\newsymbol\ngeq 2303
\newsymbol\nless 2304
\newsymbol\ngtr 2305
\newsymbol\nprec 2306
\newsymbol\nsucc 2307
\newsymbol\lneqq 2308
\newsymbol\gneqq 2309
\newsymbol\nleqslant 230A
\newsymbol\ngeqslant 230B
\newsymbol\lneq 230C
\newsymbol\gneq 230D
\newsymbol\npreceq 230E
\newsymbol\nsucceq 230F
\newsymbol\precnsim 2310
\newsymbol\succnsim 2311
\newsymbol\lnsim 2312
\newsymbol\gnsim 2313
\newsymbol\nleqq 2314
\newsymbol\ngeqq 2315
\newsymbol\precneqq 2316
\newsymbol\succneqq 2317
\newsymbol\precnapprox 2318
\newsymbol\succnapprox 2319
\newsymbol\lnapprox 231A
\newsymbol\gnapprox 231B
\newsymbol\nsim 231C
\newsymbol\ncong 231D
\newsymbol\diagup 201E
\newsymbol\diagdown 201F
\newsymbol\varsubsetneq 2320
\newsymbol\varsupsetneq 2321
\newsymbol\nsubseteqq 2322
\newsymbol\nsupseteqq 2323
\newsymbol\subsetneqq 2324
\newsymbol\supsetneqq 2325
\newsymbol\varsubsetneqq 2326
\newsymbol\varsupsetneqq 2327
\newsymbol\subsetneq 2328
\newsymbol\supsetneq 2329
\newsymbol\nsubseteq 232A
\newsymbol\nsupseteq 232B
\newsymbol\nparallel 232C
\newsymbol\nmid 232D
\newsymbol\nshortmid 232E
\newsymbol\nshortparallel 232F
\newsymbol\nvdash 2330
\newsymbol\nVdash 2331
\newsymbol\nvDash 2332
\newsymbol\nVDash 2333
\newsymbol\ntrianglerighteq 2334
\newsymbol\ntrianglelefteq 2335
\newsymbol\ntriangleleft 2336
\newsymbol\ntriangleright 2337
\newsymbol\nleftarrow 2338
\newsymbol\nrightarrow 2339
\newsymbol\nLeftarrow 233A
\newsymbol\nRightarrow 233B
\newsymbol\nLeftrightarrow 233C
\newsymbol\nleftrightarrow 233D
\newsymbol\divideontimes 223E
\newsymbol\varnothing 203F
\newsymbol\nexists 2040
\newsymbol\Finv 2060
\newsymbol\Game 2061
\newsymbol\mho 2066
\newsymbol\eth 2067
\newsymbol\eqsim 2368
\newsymbol\beth 2069
\newsymbol\gimel 206A
\newsymbol\daleth 206B
\newsymbol\lessdot 236C
\newsymbol\gtrdot 236D
\newsymbol\ltimes 226E
\newsymbol\rtimes 226F
\newsymbol\shortmid 2370
\newsymbol\shortparallel 2371
\newsymbol\smallsetminus 2272
\newsymbol\thicksim 2373
\newsymbol\thickapprox 2374
\newsymbol\approxeq 2375
\newsymbol\succapprox 2376
\newsymbol\precapprox 2377
\newsymbol\curvearrowleft 2378
\newsymbol\curvearrowright 2379
\newsymbol\digamma 207A
\newsymbol\varkappa 207B
\newsymbol\Bbbk 207C
\newsymbol\hslash 207D
\undefine\hbar
\newsymbol\hbar 207E
\newsymbol\backepsilon 237F

%
\newcount\marknumber	\marknumber=1
\newcount\countdp \newcount\countwd \newcount\countht
%
%
\ifx\pdfoutput\undefined
\def\rgboo#1{}
\def\postscript#1{\special{" #1}}		
\postscript{
	/bd {bind def} bind def
	/fsd {findfont exch scalefont def} bd
	/sms {setfont moveto show} bd
	/ms {moveto show} bd
	/pdfmark where		
	{pop} {userdict /pdfmark /cleartomark load put} ifelse
	[ /PageMode /UseOutlines		
	/DOCVIEW pdfmark}
\def\bookmark#1#2{\postscript{		
	[ /Dest /MyDest\the\marknumber /View [ /XYZ null null null ] /DEST pdfmark
	[ /Title (#2) /Count #1 /Dest /MyDest\the\marknumber /OUT pdfmark}%
	\advance\marknumber by1}
\def\pdfclink#1#2#3{%
	\hskip-.25em\setbox0=\hbox{#2}%
		\countdp=\dp0 \countwd=\wd0 \countht=\ht0%
		\divide\countdp by65536 \divide\countwd by65536%
			\divide\countht by65536%
		\advance\countdp by1 \advance\countwd by1%
			\advance\countht by1%
		\def\linkdp{\the\countdp} \def\linkwd{\the\countwd}%
			\def\linkht{\the\countht}%
	\postscript{
		[ /Rect [ -1.5 -\linkdp.0 0\linkwd.0 0\linkht.5 ]
		/Border [ 0 0 0 ]
		/Action << /Subtype /URI /URI (#3) >>
		/Subtype /Link
		/ANN pdfmark}{\rgb{#1}{#2}}}
%
%
\else
\def\rgboo#1{\pdfliteral{#1 rg #1 RG}}
\pdfcatalog{/PageMode /UseOutlines}		
\def\bookmark#1#2{
	\pdfdest num \marknumber xyz
	\pdfoutline goto num \marknumber count #1 {#2}
	\advance\marknumber by1}
\def\pdfklink#1#2{%
	\noindent\pdfstartlink user
		{/Subtype /Link
		/Border [ 0 0 0 ]
		/A << /S /URI /URI (#2) >>}{\rgb{1 0 0}{#1}}%
	\pdfendlink}
\fi

\def\rgbo#1#2{\rgboo{#1}#2\rgboo{0 0 0}}
\def\rgb#1#2{\mark{#1}\rgbo{#1}{#2}\mark{0 0 0}}
\def\pdfklink#1#2{\pdfclink{1 0 0}{#1}{#2}}
\def\pdflink#1{\pdfklink{#1}{#1}}
%
%
\newcount\seccount  
\newcount\subcount  
\newcount\clmcount  
\newcount\equcount  
\newcount\refcount  
\newcount\demcount  
\newcount\execount  
\newcount\procount  
\seccount=0
\equcount=1
\clmcount=1
\subcount=1
\refcount=1
\demcount=0
\execount=0
\procount=0
%
\def\proof{\medskip\noindent{\bf Proof.\ }}
\def\proofof(#1){\medskip\noindent{\bf Proof of \csname c#1\endcsname.\ }}
\def\qed{\hfill{\sevenbf QED}\par\medskip}
\def\references{\bigskip\noindent\hbox{\bf References}\medskip
                \ifx\pdflink\undefined\else\bookmark{0}{References}\fi}
\def\addref#1{\expandafter\xdef\csname r#1\endcsname{\number\refcount}
    \global\advance\refcount by 1}

\def\nextremark #1\par{\item{$\circ$} #1}
\def\firstremark #1\par{\bigskip\noindent{\bf Remarks.}
     \smallskip\nextremark #1\par}
\def\abstract#1\par{{\baselineskip=10pt
    \eightpoint\narrower\noindent{\eightbf Abstract.} #1\par}}
%
\def\equtag#1{\expandafter\xdef\csname e#1\endcsname{(\number\seccount.\number\equcount)}
              \global\advance\equcount by 1}
\def\equation(#1){\equtag{#1}\eqno\csname e#1\endcsname}
\def\equ(#1){\hskip-0.03em\csname e#1\endcsname}
%
\def\clmtag#1#2{\expandafter\xdef\csname cn#2\endcsname{\number\seccount.\number\clmcount}
                \expandafter\xdef\csname c#2\endcsname{#1~\number\seccount.\number\clmcount}
                \global\advance\clmcount by 1}
\def\claim #1(#2) #3\par{\clmtag{#1}{#2}
    \vskip.1in\medbreak\noindent
    {\bf \csname c#2\endcsname .\ }{\sl #3}\par
    \ifdim\lastskip<\medskipamount
    \removelastskip\penalty55\medskip\fi}
\def\clm(#1){\csname c#1\endcsname}
\def\clmno(#1){\csname cn#1\endcsname}
%
\def\sectag#1{\global\advance\seccount by 1
              \expandafter\xdef\csname sectionname\endcsname{\number\seccount. #1}
              \equcount=1 \clmcount=1 \subcount=1 \execount=0 \procount=0}
\def\section#1\par{\vskip0pt plus.1\vsize\penalty-40
    \vskip0pt plus -.1\vsize\bigskip\bigskip
    \sectag{#1}
    \message{\sectionname}\leftline{\twelvebf\sectionname} 
    \nobreak\smallskip\noindent
    \ifx\pdflink\undefined
    \else
      \bookmark{0}{\sectionname}
    \fi}
%
\def\subtag#1{\expandafter\xdef\csname subsectionname\endcsname{\number\seccount.\number\subcount. #1}
              \global\advance\subcount by 1}
\def\subsection#1\par{\vskip0pt plus.05\vsize\penalty-20
    \vskip0pt plus -.05\vsize\medskip\medskip
    \subtag{#1}
    \message{\subsectionname}\leftline{\tenbf\subsectionname}
    \nobreak\smallskip\noindent
    \ifx\pdflink\undefined
    \else
      \bookmark{0}{.... \subsectionname}  
    \fi}
%
\def\demtag#1#2{\global\advance\demcount by 1
              \expandafter\xdef\csname de#2\endcsname{#1~\number\demcount}}
\def\demo #1(#2) #3\par{
  \demtag{#1}{#2}
  \vskip.1in\medbreak\noindent
  {\bf #1 \number\demcount.\enspace}
  {\rm #3}\par
  \ifdim\lastskip<\medskipamount
  \removelastskip\penalty55\medskip\fi}
\def\dem(#1){\csname de#1\endcsname}
%
\def\exetag#1{\global\advance\execount by 1
              \expandafter\xdef\csname ex#1\endcsname{Exercise~\number\seccount.\number\execount}}
\def\exercise(#1) #2\par{
  \exetag{#1}
  \vskip.1in\medbreak\noindent
  {\bf Exercise \number\execount.}
  {\rm #2}\par
  \ifdim\lastskip<\medskipamount
  \removelastskip\penalty55\medskip\fi}
\def\exe(#1){\csname ex#1\endcsname}
%
\def\protag#1{\global\advance\procount by 1
              \expandafter\xdef\csname pr#1\endcsname{\number\seccount.\number\procount}}
\def\problem(#1) #2\par{
  \ifnum\procount=0
    \parskip=6pt
    \vbox{\bigskip\centerline{\bf Problems \number\seccount}\nobreak\medskip}
  \fi
  \protag{#1}
  \item{\number\procount.} #2}
\def\pro(#1){Problem \csname pr#1\endcsname}
%
%
%
\def\rightheadline{\hfil}
\def\leftheadline{\sevenrm\hfil HANS KOCH\hfil}
\headline={\ifnum\pageno=\firstpage\hfil\else
\ifodd\pageno{{\fiverm\rightheadline}\number\pageno}
\else{\number\pageno\fiverm\leftheadline}\fi\fi}
\footline={\ifnum\pageno=\firstpage\hss\tenrm\folio\hss\else\hss\fi}

\let\cl=\centerline

\let\sss=\scriptscriptstyle

\def\AA{{\cal A}}
\def\BB{{\cal B}}
\def\CC{{\cal C}}
\def\DD{{\cal D}}

\def\FF{{\cal F}}

\def\HH{{\cal H}}

\def\OO{{\cal O}}

\def\RR{{\cal R}}
\def\SS{{\cal S}}

\def\UU{{\cal U}}
\def\VV{{\cal V}}

\def\det{\mathop{\rm det}\nolimits}

%
\newfam\dsfam
\def\mathds #1{{\fam\dsfam\tends #1}}

\font\tends=dsrom10
\font\eightds=dsrom8
\textfont\dsfam=\tends
\scriptfont\dsfam=\eightds
%

\def\integer{{\mathds Z}}

\def\real{{\mathds R}}
\def\complex{{\mathds C}}

\def\torus{{\Bbb T}}

\def\bskip{\bigskip\noindent}

\def\bdot{\hbox{\bf .}}

\def\defeq{\mathrel{\mathop=^{\sss\rm def}}}
\def\half{{1\over 2}}

\def\quarter{{1\over 4}}
\def\thalf{{\textstyle\half}}

\def\twomat#1#2#3#4{\left[\matrix{#1&#2\cr#3&#4\cr}\right]}

%

%

%

%

\input miniltx

\ifx\pdfoutput\undefined
  \def\Gin@driver{dvips.def}  
\else
  \def\Gin@driver{pdftex.def} 
\fi

\input graphicx.sty
\resetatcatcode
%
\newdimen\savedparindent
\savedparindent=\parindent
\def\AM{{\ninerm AM~}}
\def\sAM{self-dual \AM}
\def\ssAM{{\sss{\rm AM}}}
\def\RG{{\ninerm RG~}}
\def\buA{{\hbox{\teneufm A}}}
\def\buR{{\hbox{\teneufm R}}}
%
\let\symm\circ
\def\pihalf{{\pi\over 2}}
\def\circle{{\Bbb S}}
\def\rmeven{{\rm~even}}
\def\rmodd{{\rm~odd}}
\def\mod{\mathop{\rm mod}\nolimits}
\def\gcd{\mathop{\rm gcd}\nolimits}

\def\idmat{{\bf 1}}
\def\rmSL{{\rm SL}}

\def\ssspar(#1){{\scriptscriptstyle(}#1{\scriptscriptstyle)}}
\def\sspar(#1){{\scriptstyle(}#1{\scriptstyle)}}
\def\bdot{\hbox{\bf .}}
\def\hdots{\line{\leaders\hbox to 0.5em{\hss .\hss}\hfil}}
\def\bskip{\bigskip\noindent}
\def\sfrac#1#2{\hbox{\raise2.2pt\hbox{$\scriptstyle#1$}\hskip-1.2pt
   {$\scriptstyle/$}\hskip-0.9pt\lower2.2pt\hbox{$\scriptstyle#2$}\hskip1.0pt}}
\def\shalf{\sfrac{1}{2}}

\def\today{\ifcase\month\or
January\or February\or March\or April\or May\or June\or
July\or August\or September\or October\or November\or December\fi
\space\number\day, \number\year}
\addref{Lerch}
\addref{Ostr}
\addref{Harp}
\addref{Sur}
\addref{Sos}
\addref{Swi}
\addref{Sudler}
\addref{Hof}
\addref{Beck}
\addref{Lastii}
\addref{WieZa}
\addref{HKW}
\addref{Lubi}
\addref{AFH}
\addref{BFZ}
\addref{AvKri}
\addref{HaWr}
\addref{AvJii}
\addref{KniTa}
\addref{VerMes}
\addref{ALPET}
\addref{GreNeu}
\addref{KochAM}
\addref{Thus}
\addref{KKi}
\addref{KKii}
\def\leftheadline{\sixrm\hfil Hans Koch\hfil\today}
\def\rightheadline{\sevenrm\hfil trigonometric skew-products\hfil}
%
\cl{{\twelvebf On trigonometric skew-products over irrational circle-rotations}}
\bigskip

\cl{
Hans Koch
\footnote{$^1$}
{\eightpoint\hskip-2.7em
Department of Mathematics, The University of Texas at Austin,
Austin, TX 78712.}
}

\bigskip
\abstract
We describe some asymptotic properties of trigonometric
skew-product maps over irrational rotations of the circle.
The limits are controlled using renormalization.
The maps considered here arise in connection with
the self-dual Hofstadter Hamiltonian at energy zero.
They are analogous to the almost Mathieu maps,
but the factors commute.
This allows us to construct periodic orbits under renormalization,
for every quadratic irrational,
and to prove that the map-pairs arising from the Hofstadter model
are attracted to these periodic orbits.
Analogous results are believed to be true
for the self-dual almost Mathieu maps,
but they seem presently beyond reach.

\section Introduction and main results

The work presented here was motivated in part by the difficulty
in controlling limits of products
$$
A^{\ast q}(x)\,\defeq\,
A(x+(q-1)\alpha)\cdots A(x+2\alpha)A(x+\alpha)A(x)\,,
\equation(Aastqx)
$$
when $A$ is a self-dual almost Mathieu ({\ninerm AM}) factor.
To be more precise, an \AM factor is a matrix-valued
function $A=\bigl[{E-V~-1\atop 1~~~~~0}\bigr]$
with $V(x)=2\lambda\cos(2\pi(x+\xi))$ and $E,\xi\in\real$.
The associated products \equ(Aastqx) describe
generalized eigenfunction of the Hofstadter Hamiltonian
(defined below) if $E$ belongs to its spectrum.
Interestingly, a simpler description is possible in the self-dual case
$\lambda=1$ and for energy $E=0$ [\rWieZa,\rHKW].
As we will explain below,
this leads to products \equ(Aastqx)
where $A$ is one of the two scalar functions
$$
A^s(x)={\sin(\pi(x+\sfrac{\alpha}{4}))\over
\sin(\pi(-x-\sfrac{3\alpha}{4}))}\,,\qquad
A^c(x)={\cos(\pi(x+\sfrac{\alpha}{4}))\over
\cos(\pi(-x-\sfrac{3\alpha}{4}))}\,.
\equation(sincosAx)
$$
We restrict our analysis to quadratic irrationals $\alpha$.
Taking $q\to\infty$ in \equ(Aastqx) along a sequence
of continued fractions denominators for $\alpha$,
and scaling $x$ appropriately,
we prove the existence of a nontrivial limiting function $A_\ast$.

Analogous results are believed to be true for the \sAM factors as well,
but they seem presently beyond reach.
Thus, it is useful to first study the simpler factors \equ(sincosAx).
But the methods and results described here
should be of independent interest.

The large $q$ behavior of products of the type \equ(Aastqx),
but with trigonometric factors like $A(x)=2\sin(\pi x)$,
has been studied in a variety of different contexts
[\rSudler,\rBeck,\rLubi,\rKniTa,\rVerMes,\rALPET,\rGreNeu].
The work in [\rKniTa,\rVerMes,\rGreNeu] was motivated
by questions in mathematical physics and dynamical systems as well.
Just like the work presented here,
it focuses on quadratic irrationals $\alpha$
and continued fractions denominators $q$.
Limits are considered for specific and/or individual values of $x$;
while here, we consider convergence as meromorphic functions.

We start by describing how these products are obtained from the Hofstadter model.
The Hofstadter Hamiltonian on $\ell^2\bigl(\integer^2\bigr)$
describes electrons moving on $\integer^2$,
under the influence of a flux $2\pi\alpha$ per unit cell [\rHarp,\rHof].
It is given by
$$
H^\alpha=U+U^\ast+\lambda(V+V^\ast)\,,\qquad
UVU^{-1}V^{-1}=e^{-2\pi i\alpha}\,,
\equation(HofHam)
$$
where $\lambda$ is a positive constant
and $U,V$ are magnetic translations.
We consider the Landau gauge, where
$(U\phi)(n,m)=\phi(n-1,m)$ and $(V\phi)(n,m)=e^{2\pi in\alpha}\phi(n,m-1)$.

We are interested mainly in the self-dual case $\lambda=1$.
In this case, and for irrational values of $\alpha$, the spectrum of $H^\alpha$
is a Cantor set [\rAvJii] of measure zero [\rLastii].
The metric structure of this spectrum and related quantities
result from a rich interplay between arithmetic and analysis.
The non-commuting property of the pair $(U,V)$
enters the arithmetic part via the parameter $\alpha$.
In addition, it complicates the analysis.
The following shows that this can be avoided for zero energy.
After describing the steps that lead to this simplification,
we will proceed by analyzing the resulting products \equ(Aastqx).

The operators $U$, $V$, and $H^\alpha$
commute with the two dual magnetic translations
$(\UU\phi)(n,m)=\phi(n,m-1)$ and $(\VV\phi)(n,m)=e^{2\pi im\alpha}\phi(n-1,m)$.
Restricting the Hamiltonian $H^\alpha$ to generalized eigenvectors
$\phi_\xi(n,m)=e^{-2\pi in\xi}u_n$ of the translation $\UU$,
one ends up with the \AM operator,
which leads to the above-mentioned \AM matrices.

Following [\rHKW], we consider instead generalized
eigenvectors of the diagonal translation $\UU^{-1}\VV$.
They are of the form $\psi_\xi=\Theta_\xi w$
for some sequence $w:\integer\to\complex$, where
$$
(\Theta_\xi w)(n,m)=\Theta_{\xi,n,m}w_{n+m}\,,\qquad
\Theta_{\xi,n,m}=e^{-\pi im(m+1)\alpha} e^{\pi i(m-n)\xi}\theta_{n+m}\,,
\equation(PsiThetaw)
$$
and where each $\theta_k$ can be an arbitrary phase factor.
The corresponding eigenvalue is $e^{2\pi i\xi}$.

Let us restrict now to the self-dual case $\lambda=1$.
Choosing $\theta_k=e^{\pihalf ik(k+1)\alpha}$,
a tedious but straightforward computation shows that
$H^\alpha\Theta_\xi=\Theta_\xi\HH^\alpha$, where
$$
(\HH^\alpha w)_k=
2\cos(\pi((k+1)\alpha-\xi))w_{k+1}
+2\cos(\pi(k\alpha-\xi))w_{k-1}\,.
\equation(cosHam)
$$
This defines a self-adjoint operator $\HH^\alpha$ on $\ell^2(\integer)$.
Clearly, the spectrum of $\HH^\alpha$
agrees with the spectrum of the self-dual Hofstadter Hamiltonian $H^\alpha$.
What is particularly convenient about this operator is that,
for energy $E=0$, the equation $\HH^\alpha w=Ew$
represents a one-step recursion instead of a two-step recursion:
$$
w_{k+1}=-{a_k\over a_{k+1}}\,w_{k-1}\,,\qquad
a_n=\cos(\pi(n\alpha-\xi))\,.
\equation(cosRecEo)
$$
(A one-step recursion is obtained for $\lambda\ne 1$ as well,
but its structure is less useful.)
In other words, the product setting is now commutative.

A variant of this equation has been studied in [\rHKW]
for rational values of $\alpha$.
In order to summarize some of the features,
let $\alpha=p/q$ with $q$ odd and $\gcd(p,q)=1$.
Define $\omega_k=a_1\cdots a_{k-3}a_{k-1}a_{k+2}a_{k+4}\cdots a_{2q}$
for $k=0,2,\ldots,2q$.
In particular,
$\omega_0=a_2a_4\cdots a_{2q}$ and $\omega_{2q}=a_1a_3\cdots a_{2q-1}$.
Notice that $\omega_0=\omega_{2d}$.
Thus, $\omega$ extends to a $q$-periodic sequence defined on $2\integer$.
Setting $\omega_k=\omega_{k-q}$ for all odd $k$
yields a $q$-periodic sequence on $\integer$.
The sequence $w$ defined by $w_k=i^k\omega_k$ is a solution of \equ(cosRecEo).
Notice that $w$ is a holomorphic function of the variable $\xi$.
An associated meromorphic solution
is given by $k\mapsto 1/w_{-k}$.

\smallskip
Consider now irrational values of $\alpha$.
In the absence of periodicity considerations,
the equation \equ(cosRecEo) can be solved independently
on the even and odd sublattice of $\integer$.
Thus, we restrict now to the even sublattice and write $w_{2j}=y_j$.
Let $x=\sfrac{\alpha}{4}-\sfrac{\xi}{2}$.
Setting $k=2j+1$ in \equ(cosRecEo), we obtain
the recursion $y_{j+1}=\AA(x+j\alpha)y_j$,
where $\AA=A^sA^c$ is the product
of the two functions defined in \equ(sincosAx).

Notice that $A^s$ and $A^c$ are periodic with period $1$.
By contrast, $\AA$ is periodic with periodic with $\shalf$.
A change of variables $x=x'/2$ can convert this into a period $1$,
but then $x+j\alpha$ becomes $x'+j\alpha'$, with $\alpha'=2\alpha$.
Neither a period $\shalf$ nor a new angle $\alpha'$ seems desirable.
So we propose solving \equ(cosRecEo) via two separate recursions,
$$
w_{2j}=y^s_jy^c_j\,,\qquad
y^s_{j+1}=A^s(x+j\alpha)y^s_j\,,\quad
y^c_{j+1}=A^c(x+j\alpha)y^c_j\,.
\equation(wsplit)
$$
The sequences $j\mapsto y^s_j$ and $j\mapsto y^c_j$
can be studied independently, with the two problems being very similar.
Notice that the iteration leads to products of the form \equ(Aastqx).
Our goal here is to determine the behavior of these products,
as $q\to\infty$ along a sequence of continued fractions denominators for $\alpha$,
while the argument $x$ is being rescaled appropriately.

\medskip
Consider $A\in\{A^s,A^c\}$.
Given that $A$ is periodic with period $1$,
we may assume that $\alpha$ lies between $0$ and $1$.
Let $\alpha_0=\alpha$.
The continued fractions expansion of $\alpha$
is defined inductively by setting
$c_k=\lfloor\alpha_k^{-1}\rfloor$
and $\alpha_{k+1}=\alpha_k^{-1}-c_k$ for $k=0,1,2,\ldots$.
Here $\lfloor s\rfloor$ denotes the integer part
of a real number $s$.
The best rational approximants of $\alpha$ are the rationals $p_k/q_k$
defined recursively via $p_{k+1}=c_kp_k+p_{k-1}$
and $q_{k+1}=c_kq_k+q_{k-1}$,
starting with $p_0=0$, $p_1=q_0=1$, and $q_1=c_0$.
The difference $\alpha-p_k/q_k$ can be estimated
by using the standard identity
$$
(-1)^k(q_k\alpha-p_k)=\bar\alpha_{k+1}\defeq\alpha_0\alpha_1\cdots\alpha_k\,.
\equation(baralphaDef)
$$
A non-rational $\alpha$ is said to be a quadratic irrational,
if $\alpha$ is the root of a quadratic polynomial
with integer coefficients.
A well-know fact about quadratic irrationals is that
their continued fractions expansion $\alpha=1/(c_0+1/(c_1+1/(c_2+\ldots)))$
is eventually periodic.
That is, there exist $l>0$ such that $c_{k+l}=c_k$ for sufficiently large $k$.
For this and other basic facts about continued fractions
we refer to [\rHaWr].
If $c_{k+l}=c_k$ holds for all $k\ge 0$,
then we will call $\alpha$ a periodic irrational
with period $l$.

\demo Convention(FixAlpha)
The function $A^s$ and $A^c$ defined by \equ(sincosAx) will be referred to
as the sine factor and cosine factor, respectively.
Both will also be called trigonometric factors.
Unless stated otherwise, $\alpha$ is now assumed to be
a quadratic irrational between $0$ and $1$.
Its shortest eventual period will be denoted by $l$.

Given a pair $(\alpha,A)$ with $A$ meromorphic,
and a positive integer $q$,
define $A^{\ast q}$ as in equation \equ(Aastqx).
We formulate our main result in two parts.
The first part is the following.

\claim Theorem(ProdConv)
Given a quadratic irrational $\alpha$,
there exists an even integer $n\in\{l,2l,3l,\ldots\}$
and a nonnegative integer $\mu$, such that the following holds.
$\alpha_\mu$ is periodic, and $\mu=0$ if $\alpha$ is periodic.
Let $A$ be one of the trigonometric factors defined in \equ(sincosAx).
Then the limit
$$
A_\ast=\lim_{t\to\infty}A^{\ast q_{\mu+tn}}(\bar\alpha_{\mu+tn}\,\bdot)
\equation(AProdConv)
$$
exists as an analytic function from $\complex$
to the Riemann sphere $\complex\cup\{\infty\}$,
with the convergence being uniform on compact subsets of $\complex$.
The zeros and poles of $A_\ast$ are all simple
and lie in $\real\setminus\integer[\alpha]$.

Other features of the limit \equ(AProdConv) will be described later,
after we have prepared the proper context.

One of the consequences of \clm(ProdConv) is the existence of recurrent orbits
for the recursion \equ(wsplit).
These orbits do not belong to $\ell^2(\integer)$.
But after a suitable truncation, they yield
approximate eigenfunction of $\HH^\alpha$,
and thus of $H^\alpha$ via \equ(PsiThetaw).
To be more precise, consider the one-sided sequence $y^s$
defined by setting $y^s_{j+1}=A^s(j\alpha)y^s_j$ for $j\ge 0$,
starting with $y^s_0=1$.
By \equ(AProdConv), this sequence is recurrent in the sense that
$y^s_{q_{\mu+tn}}\!\!\to A^s_\ast(0)$ as $t\to\infty$.
The same holds for the sequence $y^c$ defined via the factor $A^c$.
As an immediate consequence we have the following.
Let $y=y^sy^c$.

\claim Corollary(GenEigen)
With $y$ as defined above, and for $t\ge 0$,
define $\phi_t\in\ell^2\bigl(\integer^2\bigr)$
by setting $\phi_t(n,m)=\Theta_{\alpha,n,m}y_{(n+m)/2}$
whenever $n\pm m=0,2,4,\ldots,2q_{\mu+tn}$,
and $\phi_t(n,m)=0$ otherwise.
Then $\|H^\alpha\phi_t\|_{\ell^2}\le\epsilon_t\|\phi_t\|_{\ell^2}$
as $t\to\infty$, with $0<\epsilon_t=\OO(1/t)$.

The point here is not that
zero belongs to the spectrum of $H^\alpha$; this fact is well-known.
The interesting part of this corollary
is the information about approximate eigenvectors.
It should be noted, however, that this only involves
the convergence of \equ(AProdConv) at the origin.
The observed behavior of the sequence $j\mapsto y^s_j$ for the
inverse golden mean $\alpha=\bigl(\sqrt{5}-1\bigr)/2$
is shown in Figure 1 and described in Section 2.

A result analogous to \clm(GenEigen) was given in [\rKochAM]
for the \sAM factors;
but it required an assumption on the sequence $q\mapsto A^{\ast q}$,
as a substitute for \equ(AProdConv).

\smallskip
Consider the circles $\circle=\real\cup\{\infty\}$ and  $\torus=\real/\integer$.
Given a continuous function $A:\real\mapsto\circle$,
the recursion $y_{j+1}=A(x+j\alpha)y_j$
can be combined with a translation $x\mapsto x+\alpha$ of the line $X=\real$
to define a skew-product map $G$,
$$
G(x,y)=(x+\alpha,A(x)y)\,,\qquad
x\in X\,,\quad y\in\real\,.
\equation(GxDef)
$$
The function $A$ will be referred to as the factor of $G$.
We also use the notation $G=(\alpha,A)$.
If $A$ is periodic with period $1$,
then we will also consider the circle $X=\torus$ in \equ(GxDef).
Notice that the $q$-th iterate of $G$ is given by
$$
G^q
=\bigl(q\alpha, A^{\ast q}\bigr)\,,
\equation(GqPower)
$$
with $A^{\ast q}$ as defined in \equ(Aastqx).
So \clm(ProdConv) can be interpreted as saying that
the trigonometric map $G$ converges under proper iteration and rescaling
to a map $(\alpha_\mu,A_\ast)$.
This motivates the following renormalization approach.

As is common in the renormalization of maps on spaces $\torus\times Y$,
we first generalize the notion of periodicity by considering pairs of maps.
If $A$ is periodic with period $1$, then we pair $G=(\alpha,A)$
with the map $F=(1,1)$ whose second component is the constant function $1$.
Then the periodicity of $A$ is expressed
by the property that $G$ commutes with $F$.
More generally, let us now consider pairs $(F,G)$
of maps $F=(1,B)$ and $G=(\alpha,A)$ that commute.
Then the renormalized pair is defined by the equation
$$
\buR(F,G)=\bigl(\check F,\check G\bigr)\,,\qquad
\check F=\Lambda^{-1}G\Lambda\,,\quad
\check G=\Lambda^{-1}FG^{-c}\Lambda\,,
\equation(RGDef)
$$
where $c=\lfloor\alpha^{-1}\rfloor$ and $\Lambda(x,y)=\bigl(\alpha x,y\bigr)$.
Here, $G^{-c}$ denotes the $c$-th iterate
of the inverse map $G^{-1}=\bigl(-\alpha,A(\bdot-\alpha)^{-1}\bigr)$.
The first component of $\check F$ is $1$,
and the first component of $\check G$ is $\check\alpha=\alpha^{-1}-c$.
Notice that $\alpha\mapsto\check\alpha$ is the Gauss map
that appears in the continued fractions expansion of $\alpha$.
That is, if $\alpha=1/(c_0+1/(c_1+1/(c_2+\ldots)))$,
then $c=c_0$ and $\check\alpha=1/(c_1+1/(c_2+1/(c_3+\ldots)))$.
If $\alpha$ is periodic,
then the possibility arises that some pair
$((1,B),(\alpha,A))$ is periodic under the iteration of $\buR$.
Such periodic orbits indeed exist, as the following theorem implies.

\claim Theorem(RGPeriods)
With $\mu$, $n$, and $A$ as in \clm(ProdConv),
the limit
$$
B_\ast=\lim_{t\to\infty}A^{\ast q_{\mu+tn-1}}(\bar\alpha_{\mu+tn}\,\bdot)
\equation(BProdConv)
$$
exists and shares the properties of the limit $A_\ast$
that are described in \clm(ProdConv).
The pair $(F_\ast,G_\ast)$
with $F_\ast=(1,B_\ast)$ and $G_\ast=(\alpha_\mu,A_\ast)$
commutes and is a fixed point of $\buR^n$.

We note that \equ(AProdConv) and \equ(BProdConv) are stating
that $\buR^{\mu+tn}(F,G)\to(F_\ast,G_\ast)$ as $t\to\infty$.

A result analogous to \clm(RGPeriods) is believed to hold
for skew-product maps with \sAM factors.
The renormalization operator $\buR_\ssAM$ for such maps
is defined the same way as $\buR$,
but the scaling $\Lambda$ acts on the variable $y$ as well.
Periodic orbits for $\buR_\ssAM$ that attract \sAM pairs
are expected to exist for infinitely many energies in the spectrum;
see e.g.~the discussion in [\rKKi].
But existing proofs are restricted to
the inverse golden mean $\alpha=\bigl(\sqrt{5}-1\bigr)/2$
and cover only two periods:
a period $n=3$ [\rKochAM] that appears to attract
the \sAM pair for the largest (smallest) energy in the spectrum,
and and a period $n=6$ [\rKKii] that appears to attract
the \sAM pair for energy zero.
Both proofs are based on local methods
(computer-assisted perturbation theory about an approximate solution).
Convergence result like \equ(AProdConv) and \equ(BProdConv)
are considered global in renormalization,
and they are notoriously difficult to prove.
This was one of our main motivations
for considering skew-product maps
associated with the operator \equ(cosHam).

For further information on skew-product maps
with factors in $\rmSL(2,\real)$,
as they relate to the work presented here,
we refer to [\rKKi] and references therein.
Renormalization methods for such maps
have been used also in connection with the problem
of reducibility [\rAvKri].

\smallskip
Our proof of \equ(AProdConv) and \equ(BProdConv)
is based on estimates on the zeros of the functions
on the right hand sides of these two equations.
The limit $t\to\infty$ is performed first for the zeros;
then the functions $A_\ast$ and $B_\ast$ are constructed
from the limiting set of zeros.
The zero-sets are discussed in Section 3,
while their regularity (as sequences) and convergence
is considered in Section 4.

\section Additional observations and remarks

A property that greatly simplifies the renormalization
of \AM maps [\rKochAM] is reversibility.
The same is true for the maps considered here.

\claim Definition(reversible)
Define $\SS(x,y)=(-x,y)$.
We say that the map $G=(\alpha,A)$ is reversible (with respect to $\SS$)
if $G^{-1}=\SS G\SS$.
The function $A^\symm$ defined by $A^\symm(z)=A(z-\,\sfrac{\alpha}{2})$
will be referred to as the symmetric factor of $G$.

Notice that reversibility is preserved
under composition of commuting maps.
For the factor $A$, reversibility means that $A(x-\alpha)=A(-x)^{-1}$.
The corresponding property for $A^\symm$ is simply $A^\symm(z)^{-1}=A^\symm(-z)$.
It is not hard to see that
any meromorphic function $A^\symm$ with this property
admits a representation $A^\symm(z)=a(z)/a(-z)$, with $a$ entire analytic.
In particular, $a(z)=\sin(\pi(z-\sfrac{\alpha}{4}))$
is a possible choice for the sine factor $A^s$,
and $a(z)=\cos(\pi(z-\sfrac{\alpha}{4}))$
for the cosine factor $A^c$.
We note that the reversibility of the maps considered here
is due to our choice of the variable $x=\sfrac{\alpha}{4}-\sfrac{\xi}{2}$.

\smallskip
Figure 1 shows the sequence $j\mapsto y^s_j$ described before \clm(GenEigen),
for the inverse golden mean $\alpha=\bigl(\sqrt{5}-1\bigr)/2$.
The graph on the right fits a behavior
$y^s_j\sim Y(\log j)j^\tau$ for large $j$, with $Y$ periodic and $\tau\simeq 0.86$.
The period of $Y$ appears to be $\log\bigl(\alpha^{-6}\bigr)$,
indicating that $\buR$ has a period $6$ in this case.
The same values are found for the sequence $j\mapsto y^c_j$.

\vskip 0.5cm
\hbox{\hskip3pt
\includegraphics[height=4.2cm,width=6.7cm]{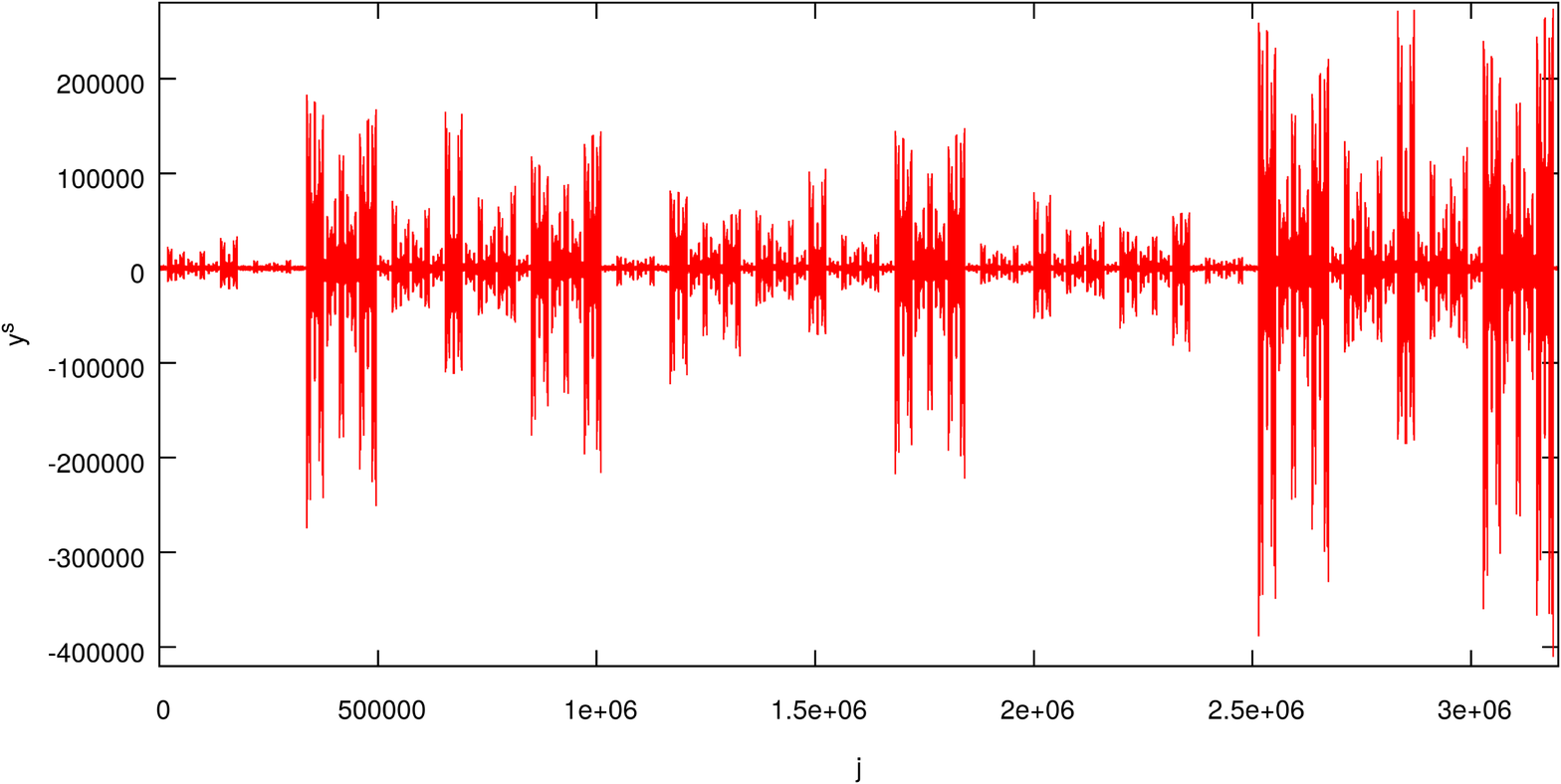}
\hskip3pt
\includegraphics[height=4.2cm,width=6.7cm]{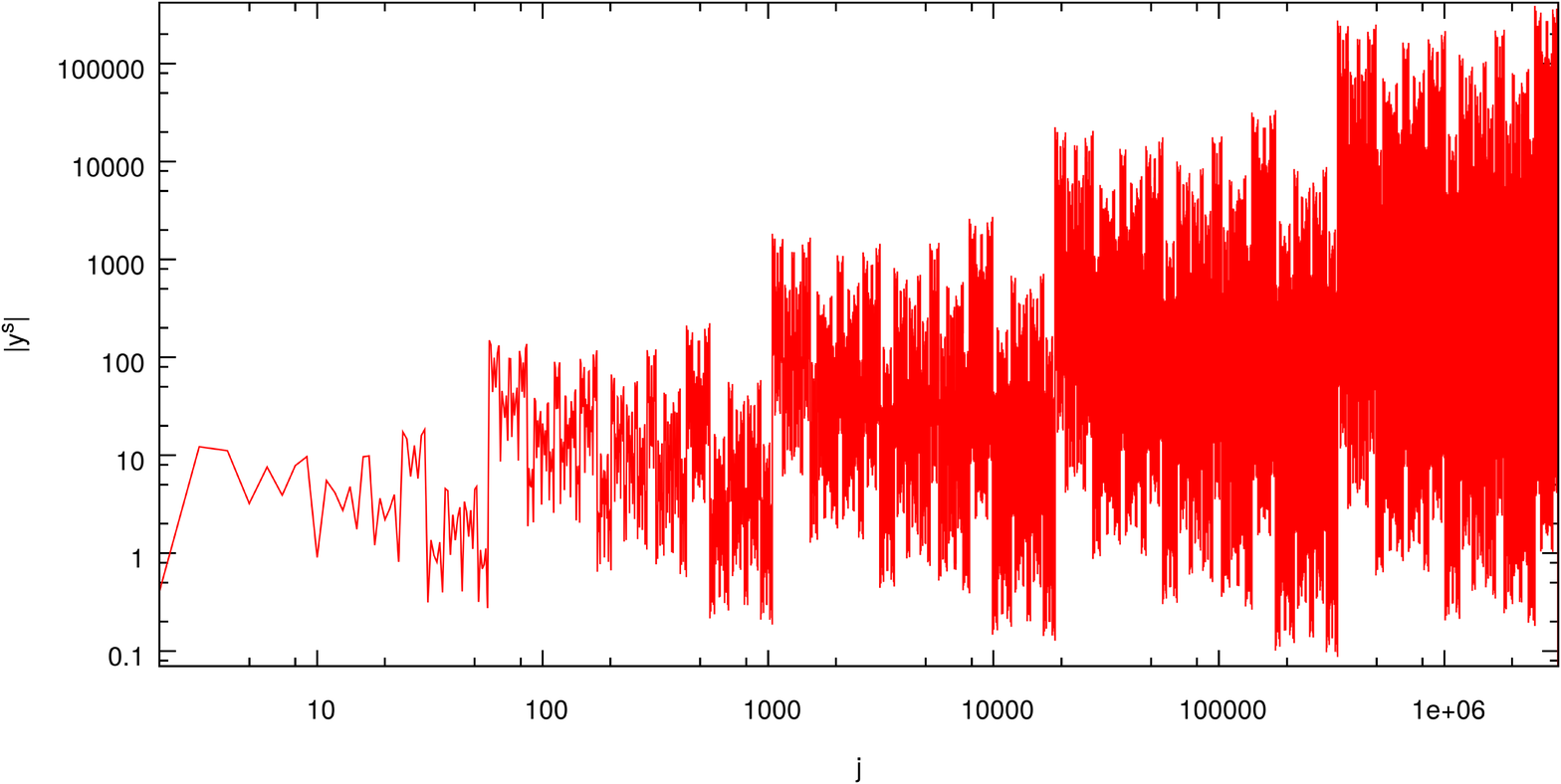}
}
\centerline{\hskip0.0cm\eightpoint{\bf Figure 1.}
The orbit $j\mapsto y^s_j$ for the inverse golden mean (left)}
\centerline{\hskip0.0cm\eightpoint
and its absolute value (log scale, right).}
\vskip 0.4cm

Similar behavior is observed for
the \sAM model at energy zero.
The main difference is a smaller growth rate, $\tau_\ssAM\simeq 0.39$.
This may be related to the fact that
the \AM operator uses (generalized eigenvectors of)
the dual vertical translation $\UU$,
while the operator $\HH^\beta$
uses the dual diagonal translation $\UU^{-1}\VV$.

\bskip
{\bf Further remarks}.

\item{$\circ$}
It should be possible to determine (maybe explicitly) minimal values
for the integers $\mu$ and $n$
that appear in Theorems \clmno(ProdConv) and \clmno(RGPeriods),
say as functions of the pre-period and period of $\alpha$.
Here, we prove little more than existence.

\item{$\circ$}
The symmetric factor $A^\circ=a/a(-\,\bdot)$
of a reversible map
is determined by the logarithmic derivative $a'/a$
of its numerator $a$.
So there is a natural additive formulation
of the methods and results described here.
See also Remark 6 in Section 4.

\item{$\circ$}
The renormalization operator $\buR$ could be defined
as a dynamical system on a suitable space of map-pairs,
but our analysis did not require this.
Our main goal here is to control the limits \equ(AProdConv) and \equ(BProdConv).

\item{$\circ$}
Consider the fixed point of $\buR^n$ described in \clm(RGPeriods),
and assume that $\buR$ has been defined properly as a dynamical system.
Based on Remark 7 in Section 4,
we expect that the derivative of $\buR^n$ at this fixed point
has an eigenvalue $\bar\sigma_n^{-1}$, where $\sigma=\alpha_\mu$.
This supports an observation made for the \sAM maps:

\item{$\circ$}
The second largest eigenvalue that was observed
numerically in [\rKKi,\rKKii] for the inverse golden mean
(with $n=6$ at energy zero)
agrees with the above-mentioned value $\bar\sigma_n^{-1}$.

\item{$\circ$}
The largest eigenvalue in the \sAM case
is believed to be associated with variations of the energy;
so we are not expecting to see it here,
where the value of the energy is fixed to zero.

\item{$\circ$}
For \sAM factors, limits like \equ(AProdConv) can exist
only if the energy $E$ is chosen in the spectrum of $H^\alpha$.
For irrational $\alpha$,
finding a point in the spectrum is nontrivial, except for $E=0$.
Given that $E=0$ also leads to a simpler equation \equ(cosRecEo),
one may wonder whether the behavior of the model near energy zero
is in some sense more trivial than near other energies.
The numerical results in [\rKKi],
which cover both zero and nonzero energies,
suggest that this is not the case.

\section Sets of zeros and renormalization

The goal in this section is to show that
the zeros of a pair ``stabilize'' under the iterating of $\buR^n$,
for a suitable choice of $n$.
By the zeros of a pair $P=(F,G)$ we mean the zeros
of the factors of $F$ and $G$.
Due to the scaling $\Lambda$,
the zeros of $P$ in a large interval
$|x|<r$ determine the zeros of $\buR^n(P)$
in an interval $|x|<cr$ that is larger by a factor $c>1$.
So it suffices to control the zeros in some fixed interval.
This is done by controlling the spacing of zeros
(via the three-gap theorem),
and by determining a pair of invariant zeros.

\smallskip
Consider the continued fractions approximants $p_k/q_k$
for a given irrational number $\alpha=1/(c_0+1/(c_1+1/(c_2+\ldots)))$
between $0$ and $1$.
The recursion relation described before \equ(baralphaDef)
can be written as
$$
\twomat{p_{k-1}}{q_{k-1}}{p_k}{q_k}=\CC_k(\alpha)\defeq
\twomat{0}{1}{1}{c_{k-1}}\cdots\twomat{0}{1}{1}{c_1}\twomat{0}{1}{1}{c_0}\,.
\equation(CFMat)
$$
Here $k\ge 1$.
$\CC_0(\alpha)$ is defined to be the identity matrix $\idmat$.
In what follows, we omit the argument $\alpha$
whenever it is clear what its values is.
Notice that $\det(\CC_k)=\pm 1$.

\claim Proposition(pnqnModFour)
Assume that $\alpha$ is periodic with period $l$.
Then there exist a positive integer $t$
such that $\CC_{tl}\equiv\idmat\,(\mod\,4)$.

\proof
Given that there are only finitely many
$2\times 2$ integer matrices modulo $4$,
we can find integers $s>r\ge 0$ such that
$\CC_l^s\equiv\CC_l^r\,(\mod\,4)$.
So the claim holds for $t=s-r$.
\qed

\claim Definition(sigmaellDef)
Let $\ell(\alpha)$ be the smallest
even value of $\ell\in\{l,2l,3l,\ldots\}$ with the property that
$\CC_{\ell}\equiv\idmat\,(\mod\,4)$.
Let $k(\alpha)$ be the smallest value of $k\ge 0$
such that $\alpha_k$ is periodic.
Define $\sigma=\alpha_k$ for $k=k(\alpha)$.

In what follows, we always consider pairs $P=(F,G)$
that commute, with $F$ of the form $F=(1,B)$.
If $m$ is even,
then the $m$-th iterate of $\buR$ can be written as
$$
\buR^m(F,G)=\bigl(\tilde F,\tilde G\bigr)\,,\qquad
\eqalign{
\tilde F&=\Lambda_m^{-1}F^{p_{m-1}}G^{-q_{m-1}}\Lambda_m\,,\cr
\tilde G&=\Lambda_m^{-1}F^{-p_m}G^{q_m}\Lambda_m\,,\cr}
\equation(tildeFtildeG)
$$
where $\Lambda_m(x,y)=(\bar\alpha_m x,y)$.
The same holds if $m$ is odd, except that the exponents
in the expressions for $\tilde F$ and $\tilde G$
have the opposite signs.

\demo Notation(sincosMap)
The map $G=(\alpha,A)$ with the sine (cosine, or trigonometric)
factor $A$ will be referred to as the sine (cosine, or trigonometric) map.

For reference later on, consider the case $B=1$.
Assume that $m$ is even.
Then, up to a scaling by $\bar\alpha_m$,
the factor of $\tilde G$ is $A^{\ast q_m}$.
Assume that $G$ is the sine map.
Regard $A$ and $A^{\ast q_m}$ as functions
on the circle $\torus=\real/\integer$.
Then $A$ has a single zero at $-\,\sfrac{\alpha}{4}$.
So $A^{\ast q_m}$ has $q_m$ zeros.
These zeros are the first $q_m$ points on the orbit
of $-\,\sfrac{\alpha}{4}$ under repeated rotation by $\alpha$.
Thus, they are all simple.
Similarly for the poles of $A^{\ast q_m}$,
which lie on the orbit of $-\,\sfrac{3\alpha}{4}$.
Since the difference between $-\,\sfrac{3\alpha}{4}$ and $-\,\sfrac{\alpha}{4}$
does not belong to $\integer[\alpha]$,
the set of poles is disjoint from the set of zero.
Thus, no cancellations occur between zeros and poles.
The same holds for odd values of $m$ and for the cosine map.

\demo Remark(sincosAoZeros)
Modulo $1$, the symmetric factor $A^\symm$ of the sine map
has its zero at $a=\sfrac{\alpha}{4}$ and its pole at $-a$.
For the symmetric factor of the cosine map,
the zero is at $a=\sfrac{\alpha}{4}-\shalf$ and the pole at $-a$.

\claim Lemma(ZeroSet)
Let $\ell=\ell(\alpha)$ and $k=k(\alpha)$.
Then there exists a nonnegative integer $\kappa=\kappa(\alpha)$
of the from $\kappa=k+s\ell$ with $s\ge 0$,
and real numbers $a_\ast$, $b_\ast$,
such that the following holds.
Let $G=(\alpha,A)$ be the sine map and let $F=(1,1)$.
If $m=\kappa+t\ell$ with $t\ge 1$,
then the symmetric factors for the renormalized maps \equ(tildeFtildeG)
satisfy $\tilde A^\symm(a_\ast)=0$ and $\tilde B^\symm(b_\ast)=0$.
If $\alpha$ is periodic, then $\kappa=0$,
$a_\ast=\sfrac{\alpha}{4}$, and $b_\ast=-a_\ast$.

\proof
Let $\kappa=k+s\ell$ with $s\ge 0$ to be determined.
Set $m=\kappa+t\ell$,
where $t$ can be any positive integer.
Consider \equ(tildeFtildeG) with this value of $m$.

We start by determining a zero of $\tilde A^\symm$.
Notice that $\CC_m\equiv\CC_k\,(\mod\,4)$
by \clm(pnqnModFour) and by the definition of $\ell$.
Consider first the case where $k$ is even and $q_k$ odd.
Then $m$ is even and $q_m$ odd, so
$$
\tilde A^\symm(z)=\prod_{|j|<q_m/2}
A^\symm(\bar\alpha_m z-j\alpha)
\qquad\quad(m \rmeven,\; q_m \rmodd)\,.
\equation(tildeAoqkodd)
$$
Recall that $A^\symm$ has a zero at $a=\sfrac{\alpha}{4}$.
In order to find a suitable zero of $\tilde A^\symm$,
imagine that $\bar\alpha_m$ is small.
Then we need to find a value of $j$
for which $j\alpha$ is close to $\sfrac{\alpha}{4}$ modulo $1$.

For every positive integer $i$, denote by $p_i'$ ($q_i'$) the remainder
in the division of $p_i$ ($q_i$) by $4$.
Let $n=k+t\ell$.
By \equ(baralphaDef) we have
$q_n\alpha-p_n=\bar\alpha_n\alpha_n$, and thus
$$
{q_n-q_n'\over 4}\alpha+{q_n'\over 4}\alpha
-{p_n'\over 4}
\equiv\quarter\bar\alpha_n\alpha_n\qquad(\mod\,1)\,.
\equation(quarterFirst)
$$
Similarly, $q_{n-1}\alpha-p_{n-1}=-\bar\alpha_n$,
and thus
$$
{q_{n-1}-q_{n-1}'\over 4}\alpha
+{q_{n-1}'\over 4}\alpha
-{p_{n-1}'\over 4}
\equiv-\quarter\bar\alpha_n\qquad(\mod\,1)\,.
\equation(quarterSecond)
$$
Now multiply \equ(quarterFirst) by $u=p_{n-1}'$
and \equ(quarterSecond) by $v=-p_n'$.
Adding the resulting congruences yields
$$
j_n\alpha
+{d\over 4}\alpha
\equiv\quarter\bar\alpha_n(u\alpha_n-v)
\qquad(\mod\,1)\,,
\equation(quarterCombined)
$$
where $d=\det(\CC_n)=\pm 1$ and
$$
j_n=u{q_n-q_n'\over 4}
+v{q_{n-1}-q_{n-1}'\over 4}+{d'-d\over 4}\,,\qquad
d'=p_{n-1}'q_n'-p_n'q_{n-1}'\,.
\equation(hatjDef)
$$
Clearly $j_n$ is an integer.
So we have
$$
A^\symm(\bar\alpha_n z-j_nd\alpha)
=A^\symm(\bar\alpha_n(z-a)+\sfrac{\alpha}{4})\,,\qquad
a={d\over 4}(u\alpha_n-v)\,.
\equation(mainFactor)
$$
Notice that this factor vanishes for $z=a$,
since $A^\symm$ has a zero at $\sfrac{\alpha}{4}$.

Assume first that $\alpha$ is periodic.
In this case we choose $s=0$.
Then the assumption that $m$ is even and $q_m$ odd
is satisfied automatically.
In fact, $\CC_n\equiv\idmat\,(\mod\,4)$, and thus
$u=d=d'=1$ and $v=0$.
So the factor \equ(mainFactor) vanishes at $a_\ast=a=\sfrac{\alpha}{4}$.
And the same holds for $\tilde A^\symm(z)$.
Next, assume that $\alpha$ is nonperiodic.
Notice that, modulo $4$, the matrix $\CC_m$
does not depend on the value of $t$.
In particular, $u$, $v$, $d$, and $d'$ are independent of $t$.
Since $m=n+s$, we can choose $s\ge 0$ in such a way that $|j_n|<q_m/2$.
This can be done independently of $t$.
Then \equ(mainFactor) implies that
$\tilde A^\symm$ vanishes at $a_\ast=a\bar\alpha_n/\bar\alpha_m$.
Notice that $\bar\alpha_n/\bar\alpha_m=\sigma_\ell^{-s}$.

This settles the case where $k$ is even and $q_k$ odd.
If $k$ and $q_k$ are both odd,
then \equ(tildeAoqkodd) holds with $z$ replaced by $-z$
on the right hand side.
So we can repeat the above, with $a$ replaced by $-a$.

Next consider the case where both $k$ and $q_k$ are even.
Then so are $m$ and $q_m$, and $\tilde A^\symm$ is given by
$$
\tilde A^\symm(z)
=\prod_{j=-q_m/2}^{q_m/2-1}A^\symm(\bar\alpha_m z+\sfrac{\alpha}{2}+j\alpha)
\qquad\quad(m \rmeven,\; q_m \rmeven)\,.
\equation(tildeAoqkeven)
$$
Now we need $j\alpha$ close to $-\,\sfrac{\alpha}{4}$ modulo $1$,
instead of $\sfrac{\alpha}{4}$.
Adapting the above arguments to this situation is trivial.

Finally, consider the case where $k$ is odd and $q_k$ even.
Then \equ(tildeAoqkeven) holds with $z$ replaced by $-z$
on the right hand side.
Now $j_n$ is the same as in the first case,
and the argument is essentially the same.

Next, consider the zeros of $\tilde B_0$.
The expression for $\tilde B_0(z)$
is the same as that for $\tilde A_0(\pm z)$,
except that there are fewer factors in the corresponding products
\equ(tildeAoqkodd) and \equ(tildeAoqkeven).
But, as in one of the cases above,
we can choose $s>0$, if necessary,
to guarantee that the factor with $j=j_n$
appears in the given product.
In the case where $\alpha$ is periodic,
one finds that $\tilde B_0(z)$ vanishes at $-\,\sfrac{\alpha}{4}$.
This concludes the proof of \clm(ZeroSet).
\qed

\demo Remark(cosZeros)
An analogous result holds for the cosine map.
It is clear that $\kappa$ can be chosen to have the same value
for both trigonometric maps.
Similarly for the integers $\mu$ and $n$ in Lemma 3.5 below.

Given integers $\mu\ge 0$ and $n\ge 1$, to be specified later,
define
$$
(F_t,G_t)=\buR^{\mu+tn}(F,G)\,,\qquad t=0,1,2,\ldots
\equation(FtGtDef)
$$
The factors of $F_t$ and $G_t$
will be denoted by $B_t$ and $A_t$, respectively;
and the symmetric factors
by $B_t^\symm$ and $A_t^\symm$, respectively.
Denote by $\BB_t$ and $\AA_t$ the set of zeros of $B_t^\symm$
and $A_t^\symm$, respectively.
To be more precise, these are subsets of $\real$.
Notice however that $B_t^\symm$ and $A_t^\symm$ are periodic with period
$r_m=\bar\alpha_m^{-1}$, where $m=\mu+tn$.
So they define functions on the circle $\torus_m=\real/(r_m\integer)$.
The corresponding zero sets on $\torus_m$
will be denoted by $\BB_t'$ and $\AA_t'$, respectively.
In what follows, we identify a circle $\real/(r\integer)$
with the interval $I_r=[-\,\sfrac{r}{2},\sfrac{r}{2})$.
In this sense, we have
$\BB_t'=\BB_t\cap I_{r_m}$ and $\AA_t'=\AA_t\cap I_{r_m}$.

For simplicity we restrict now to the sine model
and mention the cosine model only when there is a noteworthy difference.
In what follows, we use the abbreviation
$\ell=\ell(\alpha)$ and $\kappa=\kappa(\alpha)$.

\claim Proposition(ZerosRepeat)
Consider $\mu=\kappa$ and $n=\ell$.
Given any positive real number $r$, there exists infinitely many pairs
of positive integers $t_0<t_1$, such that
$\BB_{t_1}\cap I_r=\BB_{t_0}\cap I_r$ and $\AA_{t_1}\cap I_r=\AA_{t_0}\cap I_r$.

\proof
Here we use a known relationship
between the continued fractions expansion of $\alpha$
and first-return maps under a rotation by $\alpha$ [\rAFH,\rBFZ,\rThus].
Instead of starting with a circle $(-1,0]$ modulo $1$,
it is convenient to use $(-1,\alpha]$ modulo $1+\alpha$,
where $\alpha$ is ``identified'' with $0$
by being the first image of $0$ under the rotation by $\alpha$.

To simplify notation, assume that $\alpha$ is periodic, so that $\mu=0$.
Let $t>0$ and $m=tn$.
Consider first the scaled set $\buA_t\subset\torus$,
obtained from $\AA_t'$ by scaling with a factor $\bar\alpha_m=\bar\alpha_n^t$.
The points in $\buA_t$ constitute an orbit of length $q_m$
under a translation by $\alpha$ on $\torus$.
By the three-gap theorem [\rSur,\rSos,\rSwi], this orbit
divides $\torus$ into $q_m$ arcs of at most three distinct lengths.
(If there are three, then one is the sum of the two others.)

Since $\alpha$ is periodic with period $n$,
these lengths are of the form $\delta_1\bar\alpha_m$, $\delta_2\bar\alpha_m$,
and possibly $(\delta_1+\delta_2)\bar\alpha_m$.
This follows from the fact that (up to a reflection) the first-return map
of a rotation by $\alpha_0=\alpha$ on a circle $(-1,\alpha_0]$
is a rotation by $\alpha_0\alpha_1$ on the circle
$(-\alpha_0,\alpha_0\alpha_1]$; see e.g.~Lemma 2.12 in [\rThus].
After repeating this $n$ times, the rotation is by $\bar\alpha_n\alpha_n$
and the return map is to $(-\bar\alpha_n,\bar\alpha_n\alpha_n]$.
Thus, up to a scaling factor $\pm\bar\alpha_n$,
we end up with the original rotation.
(If $\mu>0$, then the same argument can be applied
to the $\mu$-th first-return map.)

This shows that neighboring points in $\AA_t$
are separated by distances $\delta_1$, $\delta_2$,
and possibly $\delta_1+\delta_2$, that are independent of $t$.
In addition, we know from \clm(ZeroSet) that each set $\AA_t$
contains a given point $a_\ast$.
So for any fixed $r>0$,
there are only finitely many possibilities for the set $\AA_t\cap I_r$ with $t\ge 1$.
By an analogous argument,
there are only finitely many possibilities for $\BB_t\cap I_r$.
\qed

Define $\BB_t(r)=\BB_t\cap I_r$ and $\AA_t(r)=\AA_t\cap I_r$.

\claim Lemma(StableSets)
There exist integers $s\ge 0$ and $\tau\ge 1$,
as well as a real number $R>0$,
such that the following holds.
Let $\mu=\kappa+s\ell$ and $n=\tau\ell$.
Consider the definition \equ(FtGtDef) with these values of $\mu$ and $n$.
Define $R_0=2R$ and $R_t=R+\bar\sigma_n^{-1}(R_{t-1}-R)$ for $t=1,2,3,\ldots$.
Then for every $t\ge 0$,
the pair of sets $\AA_{t+1}(R_{t+1})$ and $\BB_{t+1}(R_{t+1})$
is determined by the pair of sets
$\AA_t(R_t)$ and $\BB_t(R_t)$.
Furthermore, $\AA_{t+1}(R_t)=\AA_t(R_t)$
and $\BB_{t+1}(R_t)=\BB_t(R_t)$.

\proof
With $s\ge 0$ and $\tau\ge 1$ to be determined,
let $\mu=\kappa+s\ell$ and $n=\tau\ell$.
Consider the pairs $(F_t,G_t)$ as defined by \equ(FtGtDef).
Notice that the first components of $F_t$ and $G_t$
are $1$ and $\sigma$, respectively.
Here, $\sigma$ is the periodic part of $\alpha$ described in \clm(sigmaellDef).
In what follows, $q_k$ denotes the $k$-th
continued fractions denominator for $\sigma$,
and $p_k$ denotes the corresponding numerator.

The pair $(F_t,G_t)$ can be obtained
by starting with $(F_0,G_0)$ and iterating $\buR^n$ $t$ times.
The procedure is the same at each step, so consider just $t=1$.
Since $p_n$ is even and $q_n$ odd,
the product $F_0^{-p_n}G_0^{q_n}$ can be computed inductively
by starting with $H=G_0$ and joining factors $H\mapsto KHK$
with $K\in\bigl\{F_0^{-1},G_0\bigr\}$.
Notice that, if $H=(\gamma,C)$ and $K=(\delta,D)$ are reversible,
then so is the composed map $\hat H=KHK$,
and its symmetric factor is given by
$$
\textstyle
\hat C^\symm(z)=D^\symm\bigl(z+{\gamma+\delta\over 2}\bigr)
C^\symm(z)D^\symm\bigl(z-{\gamma+\delta\over 2}\bigr)\,.
\equation(KHKsFactor)
$$
Define $J_k={p_k+q_k-1\over 2}$ for $k>0$.
The above shows that the symmetric factor of $G_1$ is of the form
$$
A_1^\symm(z)=A_0^\symm(\bar\sigma_n z)\prod_{j=1}^{J_n}
U_j^\symm(\bar\sigma_n z+u_j)U_j^\symm(\bar\sigma_n z-u_j)\,,
\equation(tildeAoProd)
$$
with $U_j^\symm\in\{A_0^\symm,B_0^\symm(-\,\bdot)\}$ and $u_j\in\real$.
An analogous expression is obtained for the
symmetric factor of $F_1$. That is,
$$
\tilde B_1^\symm(z)=B_0^\symm(\bar\sigma_n z)\prod_{j=1}^{J_{n-1}}
V_j^\symm(\bar\sigma_n z+v_j)V_j^\symm(\bar\sigma_n z-v_j)\,,
\equation(tildeBoProd)
$$
with $V_j^\symm\in\{B_0^\symm,A_0^\symm(-\,\bdot)\}$ and $v_j\in\real$.

Consider first the case $n=\ell$, meaning $\tau=1$.
We note that the translations $u_j$ in the product \equ(tildeAoProd)
depend on the order in which the factors
$A_0^\circ$ and $B_0^\circ(-\,\bdot)$ are joined to that product.
Similarly for the translations $v_j$ in the product \equ(tildeBoProd).
We fix this order once and for all.
(But we are not trying to minimize the sizes of $u_j$ and $v_j$.)

In the case $n=\tau\ell$ with $\tau>1$,
we write $\buR^n$ as the $\tau$-th iterate
of $\buR^\ell$ and use the same translations in each step.
The resulting translations $u_j$ and $v_j$ depend on $\tau$,
but only via compositions.
What is important is that there is no dependence
on the factors $A_0^\symm$ and $B_0^\symm$.

Notice that a point $a$ belongs to the zero set $\AA_1$ of $A_1^\symm$
precisely if $\bar\sigma_n a$ belongs to $\AA_0$,
or if $\bar\sigma_na\pm u_j$ belongs to $\AA_0$ or $\BB_0$,
depending on whether $U_j^\symm=A_0^\symm$ or $U_j^\symm=B_0^\symm(-\,\bdot)$,
respectively.
Thus, the set $\AA_1$ is determined from $(\BB_0,\AA_0)$
via $q_n+p_n$ affine maps $g_i(z)=g_i(0)+\bar\sigma_n^{-1}z$.
Similarly, $\BB_1$ is determined from $(\BB_0,\AA_0)$
via $q_{n-1}+p_{n-1}$ affine maps $f_i(z)=f_i(0)+\bar\sigma_n^{-1}z$.
These maps $g_i$ and $f_i$ are independent
of the functions $A_0^\symm$ and $B_0^\symm$.
Furthermore, each expands by a factor $\bar\sigma_n^{-1}>1$.

Pick $R>0$ such that $|f_i(z)|>R$ and $|g_i(z)|>R$
whenever $|z|\ge R$, for all values of $i$.
This can be done independently of the choice of $\tau$
that defines $n=\tau\ell$,
since the maps $f_i$ and $g_i$ for $\tau>1$ are compositions
of the maps $f_i$ and $g_i$ for $\tau=1$.
Nor does it depend on the choice of $s\ge 0$ that defines $\mu=\kappa+s\ell$.

Let $R_0=2R$.
Consider temporarily $s=0$, and set $r=R_0$.
Then, using one of the pairs $(t_0,t_1)$ from \clm(ZerosRepeat),
define $\tau=t_1-t_0$.

{}From now on we fix $s=t_0$.
Then $\BB_1(R_0)=\BB_0(R_0)$ and $\AA_1(R_0)=\AA_0(R_0)$.

Define a function $h$ by setting $h(r)=R+\bar\sigma_n^{-1}(r-R)$.
Then we have $|f_i(z)|>h(r)$ and $|g_i(z)|>h(r)$ whenever $|z|\ge r\ge R$.
Thus, for any given $t\ge 0$,
the pair of sets $\BB_{t+1}(h(r))$ and $\AA_{t+1}(h(r))$ is determined
by the pair of sets $\BB_t(r)$ and $\AA_t(r)$, whenever $r>R$.
Applying this with $r=R_{t-1}$ and setting $R_t=h(R_{t-1})$,
for $t=1,2,3,\ldots$, we obtain the fist claim in \clm(StableSets).

Recall that
$\BB_1(R_0)=\BB_0(R_0)$ and $\AA_1(R_0)=\AA_0(R_0)$.
Since $(\BB_0(R_0),\AA_0(R_0))$ determines $(\BB_1(R_1),\AA_1(R_1))$,
and $(\BB_1(R_0),\AA_1(R_0))$ determines $(\BB_2(R_1),\AA_2(R_1))$,
we must have
$\BB_2(R_1)=\BB_1(R_1)$ and $\AA_2(R_1)=\AA_1(R_1)$.
Iterating this argument
proves the second claim in \clm(StableSets).
\qed

\section Zero sequences and logarithmic derivatives

The main goal here is to construct the limit functions \equ(AProdConv)
and \equ(BProdConv) via their sequences of zeros.

Consider the zero-sets $\AA_t$ and $\BB_t$
for the functions $A_t^\symm$ and $B_t^\symm$, respectively.
In order to obtain estimates,
we will use that these sets are regular in a suitable sense.
In particular, they have well-defined average densities
(the number of points lying between $\pm\sfrac{r}{2}$,
divided by $r$, in the limit $r\to\infty$).
This follows from the fact that these sets are periodic with period
$\bar\alpha_m^{-1}$, where $m=\mu+tn$.
Given that the number of zeros per period is $q_m$ and $q_{m-1}$, respectively,
the average densities of $\AA_t$ and $\BB_t$ are given by
$$
\rho(\AA_t)=\bar\alpha_mq_m
=\varrho_{\sss a}+\OO\bigl(\bar\sigma_n^{2t}\bigr)\,,\qquad
\rho(\BB_t)=\bar\alpha_mq_{m-1}
=\varrho_{\sss b}+\OO\bigl(\bar\sigma_n^{2t}\bigr)\,,
\equation(AAtBBtDensity)
$$
for some positive constants $\varrho_{\sss a}$ and $\varrho_{\sss b}$.
For the estimates of $\bar\alpha_mq_m$ and $\bar\alpha_mq_{m-1}$,
we have used that the eigenvalues of $\CC_n(\sigma)$
are $\bar\sigma_n$ and $\bar\sigma_n^{-1}$ in modulus,
which is well-known and follows essentially from \equ(baralphaDef).

If $\SS$ is a discrete subset of $\real$,
unbounded below and above,
we associate with $\SS$ the increasing sequence $j\mapsto s_j$
from $\integer$ onto $\SS$
with the property that $s_0$ is the nonnegative
point in $\SS$ that is closest to zero.
The sequences associated with the zero-sets
$\AA_t$ and $\BB_t$ will be denoted by
$j\mapsto a_{t,j}$ and $j\mapsto b_{t,j}$, respectively.

\claim Proposition(PerZeroBound)
There exists $c>0$ such that the following holds.
Let $m=\mu+tn$ with $t\ge 0$.
Then the sequences $j\mapsto a_{t,j}$
and $j\mapsto b_{t,j}$ satisfy the bounds
$$
|j-\rho(\AA_t)a_{t,j}|\le c\log q_m\,,\qquad
|j-\rho(\BB_t)b_{t,j}|\le c\log q_{m-1}\,,
\qquad j\in\integer\,.
\equation(PerZeroBound)
$$

\proof
For $q>1$ consider the set $D_q=\{j\alpha: 1\le j\le q\}$
modulo $1$.
The discrepancy of $\alpha$
for a subinterval $I$ of $[0,1]$ is defined as
$$
\DD_q(I)=\bigl||I\cap D_q\bigr|-q|I|\bigr|\,.
\equation(DiscrDef)
$$
A classic result [\rLerch,\rOstr] in discrepancy theory,
concerning arbitrary irrationals $\alpha>0$ of bounded type,
asserts that there exist $c'>0$ such that
$$
\DD_q([0,x])\le c'\log q\,,\qquad 0\le x\le 1\,.
\equation(DiscrBound)
$$
An analogous bound (with $c'$ increased by a factor of $2$)
holds for intervals $I=[a,b]$,
since $\DD_q([a,b])\le\DD_q([0,a))+\DD_q([0,b])$.
This implies e.g.~that the bound \equ(DiscrBound)
generalizes
to sets $D_q=\{\vartheta+j\alpha: 1\le j\le q\}$ modulo $1$,
for arbitrary $\vartheta$.

Writing $D_q=\{\delta_1,\delta_2,\ldots,\delta_q\}$
with $0<\delta_1<\delta_2<\ldots<\delta_q\le 1$, we find that
$$
\bigl|k-q\delta_k\bigr|
=\bigl|[0,\delta_k]\cap D_q\bigr|-q\delta_k\bigr|
=\DD_q([0,\delta_k])\le c\log q\,.
\equation(PerZeroBoundOne)
$$
Let $\rho=\rho(\AA_t)$.
Setting $q=q_m=\bar\alpha_m^{-1}\rho$
and $\delta_k=\bar\alpha_m a_k'$, the above yields
$$
\bigl|k-\rho\, a_k'\bigr|\le c\log q_m\,.
\equation(PerZeroBoundTwo)
$$
This is a bound on the zeros $a_1',a_2',\ldots.a_{q_m}'$
of $A_t$ in the interval $[0,r_m]$,
where $r_m=\bar\alpha_m^{-1}$.
By choosing $\vartheta$ appropriately and
changing the index to $j=k-{q_m+1\over 2}$,
we obtain the first bound \equ(PerZeroBound) for $|j|\le{q_m-1\over 2}$.
By periodicity, this bound extends to $j\in\integer$.
The second bound in \equ(PerZeroBound) is proved analogously.
\qed

Next, we consider the limit as $t\to\infty$.
By \clm(StableSets), the limit sets
$\liminf_t\AA_t$ and $\limsup_t\AA_t$ agree,
so we can choose either one to define $\AA_\ast=\lim_t\AA_t$.
Similarly define $\BB_\ast=\lim_t\BB_t$.
Denote by $j\mapsto a_{\ast,j}$ and $j\mapsto b_{\ast,j}$
the sequences associated with the sets
$\AA_\ast$ and $\BB_\ast$, respectively.
As a consequence of the last statement in \clm(StableSets),
we have the following.

\claim Corollary(StableZeros)
There exists a positive real number $\theta<1$
such that the following holds for $t_0>0$ sufficiently large.
Let $t\ge t_0$ and $m=\mu+tn$.
Then $a_{\ast,j}=a_{t,j}$ whenever $|j|\le\theta q_m$.

This allows us to bound the zeros $a_{\ast,j}$ and $b_{\ast,j}$
by using \clm(PerZeroBound).

\claim Proposition(LimZeroBound)
There exists $C>0$, such that for all $j\in\integer$
with $|j|$ sufficiently large,
$$
|j-\rho(\AA_\ast)a_{\ast,j}|\le C\log|j|\,,\qquad
|j-\rho(\BB_\ast)b_{\ast,j}|\le C\log|j|\,.
\equation(LimZeroBound)
$$
where $\rho(\AA_\ast)=\varrho_{\sss a}$
and $\rho(\BB_\ast)=\varrho_{\sss b}$.

\proof
Define $q(t)=q_m$ with $m=\mu+tn$.
The first bound in \equ(PerZeroBound) implies that
there exists $C>0$, such that
$$
|j-\varrho_{\sss a}a_{t,j}|\le\thalf C\log q(t)\,,\qquad|j|\le q(t)\,,
\equation(LimZeroBoundOne)
$$
for all $t\ge 0$.
Here we have also used \equ(AAtBBtDensity),
together with the fact that
the sequence $k\mapsto\bar\alpha_k q_k$ is bounded.

Let $\theta,t_0>0$ be as described in \clm(StableZeros).
Consider $t_1\ge t_0$ to be specified later.
Let $j\in\integer$ such that $|j|>\theta q(t_0-1)$.
Pick the smallest value of $t>t_1$ such that $|j|\le\theta q(t)$.
Then $|j|>\theta q(t-1)$.
Using that the ratio $q(t)/q(t-1)$
is bounded by some constant $c>0$ that does not depend on $t$,
it follows that $q(t)\le\theta^{-1}c|j|$.
Substituting this bound on $q(t)$ into \equ(LimZeroBoundOne),
and using that $a_{\ast,j}=a_{t,j}$,
we obtain the first bound in \equ(LimZeroBound),
provided that $t_1$ has been chosen sufficiently large.
The second bound in \equ(LimZeroBound) is proved
analogously.
\qed

\clm(LimZeroBound)
shows that the sequences $j\mapsto a_{\ast,j}$
and $j\mapsto b_{\ast,j}$ are asymptotically regular,
in the following sense.

\claim Definition(RegularSeq)
We say that a sequence $s:\integer\to\complex$ is
asymptotically regular if there exist constants $\rho,C>0$ and $J_0>1$
such that
$$
|\rho s_j-j|\le C\log|j|\qquad{\rm whenever~}|j|\ge J_0\,.
\equation(RegularSeq)
$$

A consequence of asymptotic regularity is the following.

\claim Proposition(RegSumTail)
Let $s$ be asymptotically regular,
with constant $\rho,C,J_0$ in \equ(RegularSeq).
Let $J\ge J_0$ such that $2C\log J\le J$. Then
$$
\sum_{j>J}\left|{1\over s_j}+{1\over s_{-j}}\right|
\le 8C\rho{1+\log J\over J}\,.
\equation(RegSumTail)
$$

\proof
Write $s_j=\rho^{-1}\bigl(j+c_j\log|j|\bigr)$ with $|c_j|\le C$.
Then for $j>J$,
$$
\eqalign{
\left|{1\over s_j}+{1\over s_{-j}}\right|
&={\rho\over j}
\left|{1\over 1+c_j(\log|j|)/j}-{1\over 1-c_{-j}(\log|j|)/j}\right|
\le 8C\rho{\log j\over j^2}\,.\cr}
\equation(RegSumTailOne)
$$
The bound \equ(RegSumTail) now follows from the fact that
$$
\sum_{j>J}{\log j\over j^2}
\le\int_J^\infty{\log t\over t^2}dt={1+\log J\over J}\,.
\equation(RegSumTailYwo)
$$
\qed

The above will be used to estimate logarithmic
derivatives of symmetric factors such as $A_t^\symm$.
Given a sequence of complex numbers $j\mapsto z_j$
with $z_j^{-1}=\OO(j^{-1})$ as $j\to\pm\infty$, we define
$$
{\sum_j}'{1\over z-z_j}
\defeq\lim_{J\to\infty}\sum_{j=-J}^J{1\over z-z_j}\,,
\equation(SumPrimeDef)
$$
provided that the limit exists.
To be more precise, convergence for any given $z\in\complex$
is considered a statement about the tail of the sum,
where finitely many terms may be omitted.
Thus, if the sum \equ(SumPrimeDef) converges for $z=0$,
then it converges uniformly on compact subsets of $\complex$.
The value of \equ(SumPrimeDef) at a pole is defined to be $\infty$.

Consider now pairs of functions $(\phi,\psi)$ of the form
$$
\psi(z)={\sum_j}'{1\over z-a_j}\,,\qquad
\phi(z)={\sum_j}'{1\over z-b_j}\,,
\equation(psiphiDef)
$$
that have residue $1$ at each pole.
Setting $a_j=a_{\ast,j}$ and $b_j=b_{\ast,j}$
defines the functions $\psi_\ast=\psi$ and $\phi_\ast=\phi$, respectively.
Notice that the above sums $\Sigma_j'$ converge in this case,
uniformly on compact subsets of $\complex$,
as a result of \clm(LimZeroBound) and \clm(RegSumTail).
Similarly, setting $a_j=a_{t,j}$ and $b_j=b_{t,j}$ in \equ(psiphiDef)
defines $\psi_t=\psi$ and $\phi_t=\phi$, respectively.
Convergence in this case is guaranteed by \clm(PerZeroBound).
As will become clear later, the functions $\psi_t$ and $\phi_t$
yield the logarithmic derivatives of $A_t^\symm$ and $B_t^\symm$ via
$(A_t^\symm)'(z)/A_t^\symm(z)=\psi_t(z)+\psi_t(-z)$ and
$(B_t^\symm)'(z)/B_t^\symm(z)=\phi_t(z)+\phi_t(-z)$, respectively.

\demo Remark(psiphiRG)
It is possible to generate these functions
by using an \RG transformation $\RR$
for pairs $(f,g)$ of additive skew-product maps $f=(-1,\phi)$
and $g=(\alpha,\psi)$, where $g(x,y)=(x+\alpha,\psi(x+\sfrac{\alpha}{2})+y)$
and $f(x,y)=(x-1,\phi(x-\sfrac{1}{2})+y)$.
To be more specific,
$\RR(f,g)=(\lambda^{-1}g\lambda,\lambda^{-1}fg^c\lambda)$,
with a scaling $\lambda(x,y)=(-\alpha x,y)$,
and with $c$ as described after \equ(RGDef).
But we will not need this formulation here.

\claim Lemma(ttostarConv)
$\psi_t\to\psi_\ast$ and $\phi_t\to\phi_\ast$ as $t\to\infty$,
uniformly on compact subsets of $\complex$.

\proof
Let $R>0$ and consider the restriction of $\psi_t-\psi_\ast$
to the disk $D=\{z\in\complex: |z|\le R\}$.
With $\theta$ and $t_0$ as described in \clm(StableZeros),
restrict to $t\ge t_0$ and define $J_t=\theta q(t)$, where $q(t)=q_m$ with $m=\mu+tn$.
We also assume that $t\ge t_0$ is sufficiently large,
such that $|a_{\ast,j}|>2R$ and $|a_{t,j}|>2R$ whenever $|j|>J_t$.
Then by \clm(StableZeros) we have
$$
\psi_\ast(z)-\psi_t(z)
={\sum_{|j|>J_t}}'{1\over z-a_{\ast,j}}
-{\sum_{|j|>J_t}}'{1\over z-a_{t,j}}\,,
\equation(ttostarConvOne)
$$
up to removable singularities.

Consider first $s_j=1/(z-a_{\ast,j})$ with $|j|>J_t$ and $z\in D$.
By \clm(LimZeroBound), the sequence $j\mapsto s_j$ is asymptotically regular,
with constants that are independent of $z\in D$.
Applying \clm(RegSumTail) to this sequence,
we see that the first sum in \equ(ttostarConvOne)
tends to zero as $t\to\infty$, uniformly on the disk $|z|\le R$.

Consider now the second sum in \equ(ttostarConvOne).
Due to the first bound in \equ(LimZeroBound),
the restriction $|j|>J_t$ implies that $|j-\rho(\AA_t)a_{t,j}|\le c\log|j|$,
for some constant $c>0$ that is independent of $t$.
Let $s_j=1/(z-a_{t,j})$ with $|j|>J_t$ and $z\in D$.
Clearly, the sequence $j\mapsto s_j$ is asymptotically regular,
with constants that are independent of $z\in D$ and of $t$,
for sufficiently large $t$.
Applying \clm(RegSumTail) to this sequence,
we see that the second sum in \equ(ttostarConvOne)
tends to zero as $t\to\infty$, uniformly on the disk $|z|\le R$.

This shows that the left hand side of \equ(ttostarConvOne)
tends to zero as $t\to\infty$, uniformly on $D$.
An analogous argument shows that $\phi_\ast-\phi_t\to 0$ uniformly on $D$.
Since $R>0$ was arbitrary, the assertion follows.
\qed

Now we are ready for a
\medskip\noindent
{\bf Proof of Theorems \clmno(ProdConv) and \clmno(RGPeriods)}.
Consider the set $\FF$ of meromorphic functions
whose partial fractions expansion admits a representation \equ(SumPrimeDef),
with $z_j^{-1}=\OO(j^{-1})$ as $j\to\pm\infty$.
Recall that $\psi_\ast,\phi_\ast\in\FF$ and $\psi_t,\phi_t\in\FF$ for all $t$.
In addition, these function have residue $1$ at each pole.
Clearly, if $f$ belongs to $\FF$, then so does $f(\bdot-c)$,
as well as $f(c\,\bdot)$ if $c\ne 0$.
Furthermore, the sum of two function in $\FF$ is again a function in $\FF$.

Using that $\cot\in\FF$ by symmetry, we have
$$
{\pi\sin'(\pi(z-\sfrac{\alpha}{4}))\over\sin(\pi(z-\sfrac{\alpha}{4}))}
={\sum_j}'{1\over z-a_j}\,,\qquad
{\sin(\pi(z-\sfrac{\alpha}{4}))\over\sin(\pi(-\,\sfrac{\alpha}{4}))}
={\prod_j}'\biggl(1-{z\over a_j}\biggr)\,,\qquad
\equation(sinSymmFacProd)
$$
where $a_j=\sfrac{\alpha}{4}+j$.
The product $\Pi_j'$ is defined as a limit
analogous to the limit \equ(SumPrimeDef) defining $\Sigma_j'$.
Notice the absence of an exponential factor in this product representation.
Identities analogous to \equ(sinSymmFacProd) hold
with the cosine in place of the sine;
but to simplify the description, we consider just the sine case.

By construction, $A_t^\symm(z)=a_t(z)/a_t(-z)$,
where $a_t$ is a product of translated and scaled factors
$\sin(\pi(\bdot-\sfrac{\alpha}{4}))$.
As mentioned before \dem(sincosAoZeros),
the zeros of $a_t$ are all simple,
and there are no cancellations between zeros and poles.
Thus, we have $a_t'/a_t=\psi_t$,
and the logarithmic derivative of $A_t^\symm$ admits a representation
$$
{(A_t^\symm)'(z)\over A_t^\symm(z)}
=\psi_t(z)+\psi_t(-z)={\sum_j}'\biggl({1\over z-a_{t,j}}+{1\over z+a_{t,j}}\biggr)\,.
\equation(LogDerAto)
$$
Using that $A_t^\symm(0)=1$, this implies that
$$
A_t^\symm(z)={\prod_j}'
\bigl(1-z/a_{t,j}\bigr)\bigl(1+z/a_{t,j}\bigr)^{-1}\,.
\equation(AtoProd)
$$
Now define
$$
A_\ast^\symm(z)={\prod_j}'
\bigl(1-z/a_{\ast,j}\bigr)\bigl(1+z/a_{\ast,j}\bigr)^{-1}\,.
\equation(AstaroProd)
$$
Then the analogue of \equ(LogDerAto) holds, where ``$t$'' is replaced by ``$\ast$''.
Notice that $A_\ast^\symm(0)=1$.
Using \clm(ttostarConv), together with the fact that $A_t^\symm(0)=1$ for all $t$,
we find that $A_t^\symm\to A_\ast^\symm$,
uniformly on compact subsets of $\complex$.

With a definition of $B_\ast^\symm$ analogous to \equ(AstaroProd),
the same type of argument shows that $B_t^\symm\to B_\ast^\symm$,
uniformly on compact subsets of $\complex$.
This in turn defines the skew-product maps $F_\ast$ and $G_\ast$
described in \clm(RGPeriods).
Clearly $F_\ast$ and $G_\ast$ commute,
since $F_t$ and $G_t$ commute for each $t$.

The last claim in \clm(RGPeriods) is that $P_\ast=(F_\ast,G_\ast)$ is a fixed point of $\buR^n$.
To see why this holds, consider how $\buR^n$ acts on a pair of maps $P=(F,G)$
in terms of the associated pair of functions $(\phi,\psi)$.
Denote by $\bigl(\tilde\phi,\tilde\psi)$ the pair of functions
associated with the renormalized pair of maps $\tilde P=\buR^n(P)$.
Consider first the case where $F=F_t$ and $G=G_t$ for some $t\ge 0$.
As seen in the proof of \clm(StableSets),
there exist $q_n+p_n$ affine functions $\upsilon_i$,
with $q_n$ of them being of the form $\upsilon_i(a,b)=\upsilon_i(0,0)+a$
and the other $p_n$ of the form $\upsilon_i(a,b)=\upsilon_i(0,0)+b$,
such that
$$
\tilde\psi(z)=\bar\sigma_n{\sum_j}'\sum_i
{1\over\bar\sigma_n z-\upsilon_i(a_j,b_j)}
={\sum_j}'\sum_i
{1\over z-\bar\sigma_n^{-1}\upsilon_i(a_j,b_j)}\,.
\equation(psiPoleRecursion)
$$
Since the sums $\Sigma_i$ range over a fixed finite set,
the sums $\Sigma_j'$ in this equation converge, uniformly on compact subsets of $\complex$.
Similarly, there are $q_{n-1}+p_{n-1}$
affine functions $\nu_i$ such that
$$
\tilde\phi(z)={\sum_j}'\sum_i
{1\over z-\bar\sigma_n^{-1}\nu_i(a_j,b_j)}\,.
\equation(phiPoleRecursion)
$$
The functions $\upsilon_i$ and $\nu_i$ only depend on $\sigma$ and $n$,
so the identities \equ(psiPoleRecursion) and \equ(phiPoleRecursion)
carry over to the limit $t\to\infty$.
Consider now the limit functions $\psi=\psi_\ast$ and $\phi=\phi_\ast$.
By \clm(StableSets), the set of poles $\bar\sigma_n^{-1}\upsilon_i(a_j,b_j)$
of $\tilde\psi$ agrees with $\AA_\ast$,
and the set of poles $\bar\sigma_n^{-1}\nu_i(a_j,b_j)$ of $\tilde\phi$
agrees with $\BB_\ast$.
As a result, we have $\tilde\psi=\psi$ and $\tilde\phi=\phi$.
This in turn implies that
$\tilde A^\symm=A^\symm$ and $\tilde B^\symm=B^\symm$,
since all these symmetric factors take the value $1$ at the origin.
Thus, for $P=P_\ast$ we have $\tilde P=P$, as claimed.
\qed

\demo Remark(expDir)
For the inverse golden mean (and possibly all quadratic irrationals)
it is possible to choose the translations $u_i$
in the proof of \clm(StableSets) in such a way that
one of the zeros $a_{\ast,j}$ is a fixed point
for one of the functions $a\mapsto\bar\sigma_n^{-1}(\upsilon_i(0,0)+a)$.
Given that this function has derivative $\bar\sigma_n^{-1}$,
this suggests that $\buR^n$ expands in (at least) one direction
by a factor $\bar\sigma_n^{-1}$ at $(F_\ast,G_\ast)$.
Here we use the fact that
$A_\ast^\symm\simeq A_t^\symm$ and $B_\ast^\symm\simeq B_t^\symm$ for large $t$,
so the zeros of $A_\ast$ and $B_\ast$
can be perturbed in a way that is consistent with
commutativity and reversibility,
by perturbing the zeros of the sine factor.

\bigskip\noindent
{\bf Acknowledgments}.
The author would like to thank Sa\v sa Koci\'c
for fruitful discussions.

\bigskip
\references

{\ninepoint

\item{[\rLerch]} M.~Lerch,
{\sl Question 1547},
L'Interm\'ediaire des Math\'ematiciens {\bf 11}, 144--145 (1904).

\item{[\rOstr]} A.M.~Ostrowski,
{\sl Bemerkungen zur Theorie der Diophantischen Approximationen},
Abh.\hfil\break
Math. Semin. Univ. Hamburg {\bf 1}, 77--98 (1922).

\item{[\rHarp]} P.G.~Harper,
{\sl Single band motion of conduction electrons in a uniform magnetic field},
Proc. Phys. Soc. Lond. A {\bf 68}, 874--892 (1955).

\item{[\rSur]} J.~Sur\'anyi,
{\sl \"Uber die Anordnung der Vielfachen einer reelen Zahl {\rm mod $1$}},
Ann. Univ. Sci. Budapest,
E\"otv\"os Sect. Math. {\bf 1}, 107--111 (1958).

\item{[\rSos]} V.T.~S\'os,
{\sl On the distribution {\rm mod $1$} of the sequence $n\alpha$},
Ann. Univ. Sci. Budapest,
E\"otv\"os Sect. Math. {\bf 1}, 127--134 (1958).

\item{[\rSwi]} S.~\'Swierczkowski,
{\sl On successive settings of an arc on the circumference of a circle},
Fundamenta Mathematicae {\bf 46}, 187--189 (1959).

\item{[\rSudler]} C.~Jr.~Sudler,
{\sl An estimate for a restricted partition function},
Quart. J. Math. Oxford Ser. {\bf 15}, 1--10 (1964).

\item{[\rHof]} D.R.~Hofstadter,
{\sl Energy levels and wave functions of Bloch electrons
in rational and irrational magnetic fields},
Phys. Rev. B 14, 2239--2249 (1976).

\item{[\rBeck]} J.~Beck,
{\sl The modulus of polynomials with zeros on the unit circle:
a problem of Erdos},
Ann. Math. {\bf 134}, 609--651 (1991).

\item{[\rLastii]} Y.~Last,
{\sl Zero measure spectrum for the almost Mathieu operator},
Comm. Math. Phys. {\bf 164}, 421--432 (1994).

\item{[\rWieZa]} P.B.~Wiegmann, A.V.~Zabrodin,
{\sl Quantum group and magnetic translations.
Bethe ansatz solution for the Harper's equation},
Modern Phys. Lett. B {\bf 8}, 311--318 (1994).

\item{[\rHKW]} Y.~Hatsugai, M.~Kohmoto, Y.-S.~Wu,
{\sl Quantum group, Bethe ansatz equations, and Bloch wave functions in magnetic fields},
Phys. Rev. B {\bf 53}, 9697--9712 (1996).

\item{[\rLubi]} D.S.~Lubinsky,
{\sl The size of $(q;q)_n$ for $q$ on the unit circle},
J. Number Theory {\bf 76}, 217--247 (1999).

\item{[\rAFH]} P.~Arnoux, S.~Ferenczi, P.~Hubert,
{\sl Trajectories of rotations},
Acta Arith. {\bf 87}, 209--217 (1999)

\item{[\rBFZ]} V.~Berth\'e, S.~Ferenczi, L.Q.~Zamboni,
{\sl Interactions between dynamics, arithmetic and combinatorics:
the good, the bad, and the ugly}, in: Algebraic and topological dynamics,
Contemp. Math. {\bf 385}, 333--364;
Amer. Math. Soc., Providence, RI (2005).

\item{[\rAvKri]} A.~Avila, R.~Krikorian,
{\sl Reducibility or nonuniform hyperbolicity
for quasiperiodic Schr\"odin\-ger cocycles},
Ann. Math. {\bf 164}, 911--940 (2006).

\item{[\rHaWr]} G.H.~Hardy, E.M.~Wright,
{\sl An introduction to the theory of numbers},
sixth edition, Oxford Univ. Press, New York, 2008.

\item{[\rAvJii]} A.~Avila, S.~Jitomirskaya,
{\sl The ten martini problem},
Ann. Math. {\bf 170}, 303--342 (2009).

\item{[\rKniTa]} O.~Knill, F.~Tangerman,
{\sl Self-similarity and growth in Birkhoff sums for the golden rotation},
Nonlinearity {\bf 24}, 3115--3127 (2011).

\item{[\rVerMes]} P.~Verschueren, B.~Mestel,
{\sl Growth of the Sudler product of sines at the golden rotation number},
J. Math. Anal. Appl. {\bf 433}, 200--226 (2016).

\item{[\rALPET]} C.~Aistleitner, G.~Larcher, F.~Pillichshammer, S.S.~Eddin, R.F.~Tichy,
{\sl On Weyl products and uniform distribution modulo one},
Monatshefte f\"ur Math. {\bf 185}, 365--395 (2018).

\item{[\rGreNeu]} S.~Grepstad, M.~Neum\"uller,
{\sl Asymptotic behaviour of the Sudler product of sines for quadratic irrationals},
J. Math. Anal. Appl. {\bf 465}, 928--960 (2018).

\item{[\rKochAM]} H.~Koch,
{\sl Golden mean renormalization
for the almost Mathieu operator and related skew products},
Preprint 2019.

\item{[\rThus]} J.M.~Thuswaldner,
{\sl $S$-adic sequences. A bridge between dynamics, arithmetic, and geometry.},
Preprint 2019.

\item{[\rKKi]} H.~Koch, S.~Koci\'c,
{\sl Renormalization and universality of the Hofstadter spectrum},
Preprint 2019, to appear in Nonlinearity.

\item{[\rKKii]} H.~Koch, S.~Koci\'c,
{\sl Orbits under renormalization of skew product maps
over circle rotations}.
In preparation.

}
\bye